\documentclass[twocolumn,twocolappendixa,trackchanges]{aastex631}

\shorttitle{Outflows and jets in the outer Galaxy} 
\shortauthors{T. Ikeda et al.} 

\begin{document}

\title{The detection of spatially resolved protosteller outflows and episodic jets in the outer Galaxy} 

\correspondingauthor{Toki Ikeda} 
\email{tokiikeda0122@gmail.com} 

\author[0009-0004-5964-3892]{Toki Ikeda} 
\affiliation{Department of Natural Environmental Science, Graduate School of Science and Technology, Niigata University, Ikarashi-ninocho 8050, Nishi-ku, Niigata 950-2181, Japan}

\author[0000-0002-0095-3624]{Takashi Shimonishi} 
\affiliation{Institute of Science and Technology, Niigata University, Ikarashi-ninocho 8050, Nishi-ku, Niigata 950-2181, Japan}

\author[0000-0003-1604-9127]{Natsuko Izumi} 
\affiliation{National Astronomical Observatory of Japan, 2-21-1 Osawa, Mitaka, Tokyo 181-8588, Japan}

\author[0000-0002-2699-4862]{Hiroyuki Kaneko}
\affiliation{Environmental Science Program, Faculty of Science, Niigata University, Ikarashi-ninocho 8050, Nishi-ku, Niigata, 950-2181, Japan}

\author[0000-0002-7287-4343]{Satoko Takahashi} 
\affiliation{National Astronomical Observatory of Japan, 2-21-1 Osawa, Mitaka, Tokyo 181-8588, Japan}
\affiliation{Astronomical Science Program, The Graduate University for Advanced Studies, SOKENDAI, 2-21-1 Osawa, Mitaka, Tokyo 181-8588, Japan}

\author[0000-0002-6907-0926]{Kei E. I. Tanaka} 
\affiliation{Department of Earth and Planetary Sciences, Institute of Science Tokyo, Meguro, Tokyo, 152-8551, Japan}

\author[0000-0002-2026-8157]{Kenji Furuya} 
\affiliation{Department of Astronomy, Graduate School of Science, University of Tokyo, Tokyo 113-0033, Japan}
\affiliation{RIKEN Cluster for Pioneering Research, 2-1 Hirosawa, Wako-shi, Saitama 351-0198, Japan}

\author[0000-0003-3579-7454]{Chikako Yasui} 
\affiliation{National Astronomical Observatory of Japan, 2-21-1 Osawa, Mitaka, Tokyo 181-8588, Japan}

\begin{abstract}
We present the first detection of spatially resolved protostellar outflows and jets in the outer Galaxy.
We observed five star-forming regions in the outer Galaxy (Sh 2--283, NOMF05-16/19/23/63; galactocentric distance = 15.7--17.4 kpc) with the Atacama Large Millimeter/submillimeter Array (ALMA). 
Towards Sh 2--283, we have detected distinct outflow ($\sim$5--50 km s$^{-1}$) and jet components ($\sim$50--100 km s$^{-1}$) associated with the protostar in CO(3--2) emission.
The outflows and jets are well-collimated, with the jets exhibiting multiple bullet structures.
The position-velocity diagram along the CO flow axis shows two characteristic structures: 
(a) the flow velocity which linearly increases with the position offset from the core center (Hubble-like flow), 
and (b) continuous velocity components of the periodical flows (spine-like structures), which may indicate the episodic mass-ejection event. 
The time intervals of the mass-ejection events are estimated to be 900--4000 years based on the slopes of these spine-like structures. 
These characteristics align with those of nearby protostellar systems, indicating that early star formation in low-metallicity environments, such as the outer Galaxy, resembles that in the inner Galaxy.
In contrast to the physical similarities, the $N\mathrm{(SiO)}$/$N\mathrm{(CO)}$ ratio in the jet bullet appears to be lower than that measured in the low-mass protostellar sources in the inner Galaxy.
This may indicate the different shock chemistry or different dust composition in the outer Galaxy source, although non-LTE effects could also affect the observed low $N\mathrm{(SiO)}$/$N\mathrm{(CO)}$ ratio.
We also report the new detection of the other 4 outflow sources in the outer Galaxy. 
\end{abstract}



\section{Introduction} \label{sec_intro} 
Since its first discovery by \cite{snell1980}, protostellar outflows have been universally detected toward star-forming regions \citep[][and reference there in]{Bally2016}. 
Protostars grow through mass-accretion processes, and protostellar outflows are believed to play a crucial role in removing the excess angular momentum generated by the accretion process from the protostellar envelope.

Observations of protostellar systems suggest that mass ejection phenomena are composed of multiple velocity components: jets and outflows.
Jets ($\sim$50--200 km s$^{-1}$ relative to the systemic velocity) are often detected as episodic flows (a.k.a. bullets or knots) and show collimated structures 
\citep[opening angle of $\sim$5$^\circ$--20$^\circ$;][]{Santiago2009}.
Such high velocity jets are thought to be originated from the protostellar inner envelope and bullet structures are possibly related to the periodic mass-accretion events \citep[e.g.,][]{Matsushita2019, Dutta2022,Fischer2023, Dutta2024, Takahashi2024}.
Thus, jets would be directly related to the mass-accretion rate of the protostar and its evolution processes \citep[see, e.g., reviews by][]{Arce2007, Bally2016}.
On the other hand, outflows ($\sim$ 10--50 km s$^{-1}$) usually show continuous and wide opening angle flows 
\citep[opening angle of $\sim$30$^\circ$--60$^\circ$;][]{Santiago2009, Hirano2010}. 

Such outflow and jet structures have been detected by (sub)millimeter, infrared, and optical observations especially toward the low-mass protostellar sources \cite[e.g.,][and reference therein]{bachiller90, Bac96,Lee2000, Hirano2010, Dutta2022, Takami2023, Omura2024, Dutta2024, Pyo2024, Takahashi2024}.
For high-mass protostars, the number of samples of such jet components is still limited \citep[e.g.,][]{Qiu2009, Torrelles2011,Cheng2019, Zinchenko2020,Zinchenko2021, Zinchenko2024}.
This would be due to their short existing time in the protostellar phase, limited sensitivities, and angular resolution because of their large distance from us.



The effects of environmental factors (e.g., metallicity) on the origin and driving mechanisms of outflows and jets have not been thoroughly investigated.
Because the cosmic metallicity is increasing with the time evolution of the Universe, observing protostellar cores in low-metallicity environments is crucial to understand the star formation and chemical evolution processes in the past Milky Way or those in high-redshift galaxies \citep[e.g.,][]{Balestra2007, Rafelski2012, Delgado2019}.
Atacama Large Millimeter/submillimeter Array (ALMA) enables us to spatially resolve the protostellar cores ($\sim$0.1 pc scale) located in distant low-metallicity environments such as the Large Magellanic Cloud (LMC), the Small Magellanic Cloud (SMC), and the outer Galaxy.
Recent ALMA observations have detected massive protostellar cores associated with outflows in these regions \citep[e.g.,][]{Fukui2015, Shimonishi2016, Shimonishi2023, Tokuda2022, Tok25}. 
Those with compact (size $\lesssim$0.1 pc), dense ($n_{\rm{H_2}}$ $\gtrsim$10$^6$ cm$^3$), and high-temperature (T $\gtrsim$100 K) molecular gas (i.e. hot molecular core) are also detected \citep{Shimonishi2016, Shimonishi2020, Shimonishi2021, Shimonishi2023,  Sewilo2018, Sewilo2022,  Gol24}.
James Webb Space Telescope (JWST) is also a powerful telescope for observing outflows in distant low-metallicity environments.
Based on JWST observations, \cite{Izumi2024} detected several outflow structures in star-forming regions in the outer Galaxy with a spatial resolution of 1000 au.
These observations indicate that the physical and chemical events at the early stage of star formation, such as the launching of outflows and the emergence of hot cores, may be ubiquitous even in low-metallicity environments, although the number of samples is still limited. 

While outflows have been detected in low-matellicity environments, their jet components have not been detected so far. 
\cite{Shimonishi2021} detected high-velocity SiO lines toward a protostellar source in the outer Galaxy, WB89-789 SMM2.
It showed the symmetrical structure both in the spectra and the integrated intensity map of each red- and blue-shifted component.
However, the detailed flow structures were not spatially resolved due to the insufficient angular resolution, and also the major outflow tracer, CO, was not covered by the observations. 
Hence, there is no sample of spatially resolved protostellar jets in low-metallicity environments, highlighting the need for high-resolution studies in such regions.

In this paper, we report the first detection of spatially resolved molecular outflows including their jet components in the outer Galaxy based on ALMA observations toward the star-forming region, Sharpless 2-283 (hereafter Sh 2--283). 
In addition, we also report the detection of CO outflows in the other 4 protostellar sources in the outer Galaxy. 
In Section \ref{sec_tarobsred}, we describe the observation details and the data reduction.
The observational results and analysis of the detected outflows and jets are presented in Section \ref{sec_res}.
The morphology, dynamical parameters, driving mechanisms, and chemical properties of the outflows and jets are discussed in Section \ref{sec_dis}.
The summary of this paper is given in Section \ref{sec_sum}.

\section{Targets, Observations, and data reduction} \label{sec_tarobsred} 
\subsection{Targets} \label{sec_tar} 
The target star-forming region is Sh 2--283.
This region is known as a H$_\mathrm{II}$ region located in the outer Galaxy \citep{Fich1991}.
Based on the model A5 in \cite{Reid2014}, the kinematic distance of Sh 2--283 is estimated to be 7.2 kpc, which corresponds to the galactocentric distance ($D_\mathrm{GC}$) of 15.2 kpc. 
According to the astrometric data of a nearby cluster member (ALS 18674) in the Gaia Early Data Release 3 \citep[Gaia source ID 3119827723711464576,][]{Gaia2021,Bra19}, the distance to Sh 2--283 is estimated to be 7.9$^{+1.2}_{-1.1}$ kpc, which corresponds to the galactocentric distance of 15.7$^{+1.1}_{-0.9}$ kpc. 
The distance was directly derived from the parallax (0.1269759 mas), and the parallax bias was corrected following the method described in \cite{Lindegren2021}.
In this paper, we use the distance value derived from the Gaia data for determining dynamical properties.
The metallicity of Sh 2--283 is estimated to be approximately 30 $\%$ of the solar value based on the optical spectroscopy of the associated H$_\mathrm{II}$ region \citep{Fer17}. 
Moreover, the other 4 star-forming regions \citep[NOMF05-16, 19, 23, and 63; ][]{Nakagawa2005} were separately covered in our ALMA observations.
Although the metallicities of these sources have not been measured, they are expected to be 20-30 $\%$ of the solar value based on the relation between the galactocentric distance and metallicity \citep[e.g., ][]{Are21}.

Toward these regions, we observed in total 16 protostar candidates with ALMA. 
Information of the observed sources is summarized in Table \ref{ta_source_prop}. 
For Sh 2--283 region, protostar candidates were selected by using the near-infrared data from the UKIDSS survey of the UKIRT 3.8 m telescope \citep{Lucas2008}. 
For the other regions, a near-infrared imaging survey was conducted with the Gemini--South 8.1m telescope in Chile (Proposal ID: S22A-125, PI: N. Izumi) and IRSF 1.4m telescope in South Africa, whose details will be presented in a future paper (Izumi et al. in prep.). 
We have selected 11 reddened sources with [H] -- [K] $>$ 1.3 mag, where [H] and [K] indicate H-band (1.65 $\mu$m) and K-band (2.16 $\mu$m) magnitude. 
Such a near-infrared color is consistent with theoretically-predicted colors of embedded protostars \citep{Rob06}. 
We have also selected additional 4 sources (Sh 2--283-3, Sh 2--283-4, Sh 2--283-5, and NOMF05-19--4), whose [H] -- [K] values are less than the above threshold or very faint in K-band, but they are very bright in mid-infrared and have [K] -- [22 $\mu$m] $>$8 mag, according to the  \citep{Wri10}. 
Although WISE 22 $\mu$m data may be contaminated by nearby sources due to its low spatial-resolution (12$\arcsec$ at 22 $\mu$m), such mid-infrared bright sources would potentially indicate the presence of Class I protostars. 
In addition, we included the position of NOMF05-16–1, because high-velocity CO($J$ = 2--1) components were detected in previous ALMA mapping observations towards the whole NOMF05-16 region (Izumi et al. in prep.). 
The location and near-infrared color composite images of the targets are shown in Appendix \ref{sec_position}.

\subsection{Observations} \label{sec_obs}  
 All data used in this work were acquired by ALMA as a Cycle 9 program (2022.1.01270.S, PI: T. Shimonishi).
The band 7 receiver was employed to observe molecular emission lines (such as $^{12}$CO($J$ = 3--2), HCO$^+$($J$ = 4--3), SiO($J$ = 8--7), CH$_3$OH, SO$_2$, and SO) and the 0.87-mm continuum.
Five spectral windows (SPWs) of the correlator were used in the frequency division mode.
The channel spacing for all SPWs was set to 976.6 kHz. 
The two SPWs with a bandwidth of 0.938 GHz cover the sky frequencies of 344.038--344.976 and 344.978--345.916 GHz.
The three SPWs were set to cover the sky frequencies of 345.919--347.794, 346.169--358.044, and 356.961--358.836 GHz, in which the bandwidth was 1.875 GHz.
Five protostar candidates located in the Sh 2--283 star-forming region were separately observed in October 2022.
The antenna configuration was C-2, resulting in a minimum and maximum baseline length of 15 m and 313 m, respectively.
The on-source time per source was 27.2 minutes.
The observations for Sh 2--283 were conducted in three execution blocks.
During the observation period, 45 antennas were operated except for one execution, in which 42 antennas were used.
The bandpass and flux calibrators for protostar candidates in Sh 2--283 were J0510+1800 and J0750+1231, and the phase calibrator was J0641-0320.

NOMF05-63--1 was observed in December 2022.
The 45 antennas with the antenna configuration of C-3 (minimum and maximum baseline length of 15 m  and 500 m, respectively) were used.
The on-source time for NOMF05-63--1 was 25.8 minutes.
The bandpass/flux and phase were calibrated with J0811-4929 and J0811-4929, respectively.
The other observations toward the NOMF05-16, 19, and 23 region were carried out in December 2022 for the rest 8 protostar candidates using 46 antennas with the C-3 configuration (minimum and maximum baseline length of 15 m  and 500 m, respectively). 
The on-source time for each source was 24.3 minutes.
The bandpass/flux calibrators of J1037-2934 and J0519-4546 and the phase calibrator of J0828-3731 were used for these targets.

The observations and resultant image properties are summarized in Table \ref{tab_Obs}.

\begin{deluxetable*}{ l|c c c c c c c c c c}
\tablecaption{Source Information}
\tabletypesize{\footnotesize} 
\tablehead{ 
\colhead{Source Name} & \colhead{RA$^\mathrm{a}$} & \colhead{Dec.$^\mathrm{a}$} & \colhead{$D^\mathrm{b}$} & \colhead{ $D_\mathrm{GC}^\mathrm{b}$} & \colhead{[$K$]} & \colhead{[$H$]-[$K$]} & \colhead{$V_\mathrm{sys}^\mathrm{c}$} & \colhead{Outflow$^\mathrm{d}$} & \colhead{Submm. Cont.$^\mathrm{e}$} \\  
\colhead{}            & \colhead{(J2000.0)}       & \colhead{(J2000.0)}         & \colhead{(kpc)}          & \colhead{(kpc)}                       & \colhead{(mag)} & \colhead{(mag)}       & \colhead{(km s$^{-1}$)}               & \colhead{(Y/N)}    & \colhead{(Y/N)}
}
\startdata 
Sh 2--283--1a SMM1 & 06:38:29.66 & 0:44:40.73 & 7.9 & 15.7 &  15.6   & 1.9 &   53.2 & Y & Y \\
Sh 2--283--1b & 06:38:29.60 & 0:44:41.25 & 7.9 & 15.7&   15.9   & 1.7 & 53.2 & N & N  \\
Sh 2--283--1c  & 06:38:29.38 & 0:44:36.25 & 7.9 & 15.7 & 15.2   &1.5 & 53.2 & N & N  \\
Sh 2--283-2 & 06:38:30.14 & 0:44:57.95 & 7.9 & 15.7 &  16.4   & 1.5 & 51.9 & N & Y  \\
Sh 2--283-3 & 06:38:27.83 & 0:44:42.20 & 7.9 & 15.7 &  14.0   & 0.7$^\mathrm{f}$& 47.5  & N & Y  \\
Sh 2--283-4 & 06:38:27.83 & 0:44:42.25 & 7.9 & 15.7 &  12.7   & 0.5$^\mathrm{f}$& 47.5 & N & Y  \\
Sh 2--283-5 & 06:38:28.60 & 0:44:25.00& 7.9 & 15.7 &  14.7    & 1.1$^\mathrm{f}$& 45.4   & N & Y  \\
NOMF05-16--1 & 08:03:37.95 & -37:08:06.38 & 12.8 & 17.3 &  --   &-- & 107.4 & Y & N  \\
NOMF05-19--1 & 08:06:31.80 & -37:29:52.43 & 12.7 & 17.1 &   16.3   & 1.4 & 39.5 & N & Y  \\
NOMF05-19--2 & 08:06:33.78 & -37:30:14.27 & 12.7 & 17.1 &   14.7   & 2.9 & 38.1 & N & N   \\
NOMF05-19--3 & 08:06:33.17 & -37:29:49.40 & 12.7 & 17.1 &   15.3   & 2.3 & 35.5 & N & Y  \\
NOMF05-19--4 & 08:06:31.80 & -37:29:52.57 & 12.7 & 17.1 &  11.5    & 0.8$^\mathrm{f}$ & 36.6 & N & Y  \\
NOMF05-23--1 & 08:05:33.01 & -39:09:24.61 & 13.2 & 17.4 &   14.4   & 1.4 & 108.4 & Y  &  Y \\
NOMF05-23--2 & 08:05:33.87 & -39:09:24.97 & 13.2 & 17.4 &   14.1  &1.7 & 109.0 & N & N  \\
NOMF05-23--3 & 08:05:34.81 & -39:09:16.85 & 13.2 & 17.4  &   15.6 &1.3 & 105.7 & Y & Y \\
NOMF05-63--1 & 08:29:53.20 & -45:55:09.63 & 14.1 & 17.3 &  10.3  & 1.7 & 110.3 & Y & Y \\
\enddata
\tablecomments{
See Section \ref{sec_tar} for the details of the target selection. 
$^\mathrm{a}$The peak position of continuum emission except for Sh 2--283--1b, Sh 2--283--1c, NOMF05-16--1, NOMF05-19--2, and NOMF05-23--2. 
For Sh 2--283-5 and NOMF05-16--1, the position corresponds to the emission peak of HCO$^+$($J$ = 4--3) because of the no detection of continuum emission. 
The near-infrared source positions are tabulated for Sh 2--283--1b, Sh 2--283--1c, NOMF05-19--2, and NOMF05-23--2. 
$^\mathrm{b}$For sources located in Sh 2--283 , both $D$ and $D_\mathrm{GC}$ are derived with the Gaia Early Data Release 3 \citep[Gaia source ID83
3119827723711464576:][]{Gaia2021}.
For the other sources, we use the values derived in \cite{Nakagawa2005}.
$^\mathrm{c}$The systemic velocities are derived from the molecular lines for each source.
$^\mathrm{d}$CO(3-2) outflow detection.
$^\mathrm{e}$ 0.87-mm continuum detection. 
$^\mathrm{f}$[H] -- [K] $<$ 1.3 mag, but [K] -- [22 $\mu$m] $>$8 mag.}

\label{ta_source_prop}
\end{deluxetable*}

\begin{deluxetable*}{lcccccccccc}
\tablecaption{Summary of ALMA Observations \label{tab_Obs}} 
\tablewidth{0pt} 
\tabletypesize{\footnotesize} 
\tablehead{
\colhead{Region}   & \colhead{Observation} & \colhead{Number} & \colhead{On-Source} & \colhead{Mean}                             & \colhead{Number}   &  \multicolumn{2}{c}{Baseline\tablenotemark{b}}     &  \colhead{}                                                  &  \colhead{}                                         &  \colhead{Channel}  \\
\cline{7-8}  
\colhead{}   & \colhead{Date} & \colhead{of} &  \colhead{Time}         &  \colhead{PWV\tablenotemark{a}} & \colhead{of}              &  \colhead{L5} & \colhead{L80} & \colhead{Beam Size\tablenotemark{c}}         & \colhead{MRS\tablenotemark{d}}     &  \colhead{Spacing}  \\
\colhead{}   & \colhead{}  &  \colhead{Pointings}  &  \colhead{(min)}         &  \colhead{(mm)}                             & \colhead{Antennas} & \colhead{(m)} & \colhead{(m)}   &  \colhead{($\arcsec$ $\times$ $\arcsec$)}      & \colhead{($\arcsec$)}                       &  \colhead{(kHz)}                }
\startdata 
Sh 2--283                                   &  2022 Oct 27--28 & 5 & 27.2                           &  0.3                                         &  42--45                       &  25.1               &  132.7              &  0.79 $\times$ 0.68                                     &   6.3                              &  976.6  \\
NOMF05-16     &   2022 Dec 29  & 1   &  24.3              &    0.3                    &       46         &  33.0  &   220.6    &  0.61 $\times$  0.50                                   &      4.8                &  976.6    \\
NOMF05-19 &   2022 Dec 29 &  4   &  24.3              &    0.3                    &       46         &  33.1  &   220.8    &  0.60 $\times$  0.49                                   &      4.8                &  976.6    \\
NOMF05-23 &    2022 Dec 29  &  3  &  24.3              &    0.3                    &       46         &  32.8  &   219.0    &  0.61 $\times$  0.50                                   &      4.8                &  976.6    \\
NOMF05-63 &    2022 Oct 8  & 1   &  25.8              &    0.3                   &       45        &  34.4  &   225.6    &  0.60 $\times$  0.51                                   &      4.8                &  976.6 \\
\enddata
\tablenotetext{a}{Precipitable water vapor.}
\tablenotetext{b}{L5/L80 indicate the length that includes the 5th/80th percentile of all projected baselines.}
\tablenotetext{c}{The average beam size of line images.}
\tablenotetext{d}{Maximum Recoverable Scale.}
\end{deluxetable*}

\subsection{Data reduction} \label{sec_red} 
Data reduction, including calibration and imaging, was made with the Common Astronomy Software Applications package  \citep[CASA: ][]{CASA2022} version 6.4.1.12.
The raw data were calibrated with the pipeline script provided by the observatory for the initial data flagging, the calibration of bandpass, complex gain, flux scaling, and imaging (pipeline version 2022.2.0.64) for all sources.
The visibilities of the line emissions were made by subtracting the continuum visibilities using {\itshape uvcontsub} task in CASA.
The clean images were produced using {\itshape tclean} task in CASA. 
The line image cubes were obtained adopting Briggs weighting with a robust parameter of 0.5 for each SPW.

For Sh 2--283--1a SMM1, since a variety of molecular emission were detected (see Figure \ref{image_spec_sum}), we performed phase self-calibration using the continuum after the calibration in order to improve the signal to noise ratio.
One round of phase self-calibration was applied with the solution intervals of 60 seconds.
Finally, we corrected the primary beam patterns using {\itshape impbcor} task.

The systematic error on the absolute flux was approximately 10 \% according to the ALMA technical handbook \footnote{\url{https://almascience.nao.ac.jp/documents-and-tools/cycle11/alma-technical-handbook}}.

 \begin{figure}[th!]
\begin{center}
\includegraphics[width=8.0cm]{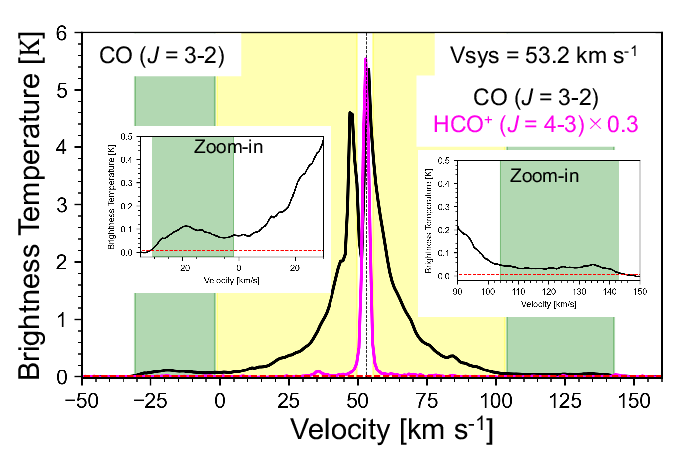}
\caption{The spectra of CO($J$ = 3--2), which is extracted from the 4$\farcs$0 $\times$ 10$\farcs$0 elliptical region around Sh 2--283--1a SMM1 (Figure \ref{image_co} (b)).
The magenta line represents the HCO$^+$($J$ = 4--3) spectra extracted from the 0$\farcs$89 $\times$ 0$\farcs$70 elliptical region at the emission peak of the 0.87-mm continuum.
The black dotted line shows the systemic velocity of the target (V$_{\rm{sys}}$ = 53.2 km s$^{-1}$).
The yellow- and green-colored areas represent the outflow and jet components, respectively.
The red dotted line represents the 3$\sigma$ rms noise level (1$\sigma$ = 0.02 K).
The smaller panels on the left and right are the zoom-in images of the CO($J$ = 3--2) spectra for the red- and blue-shifted components of the jets, respectively.}
\label{image_co_spec}
\end{center}
\end{figure}

 \begin{figure*}[tp!]
\begin{center}
\includegraphics[width=18.0cm]{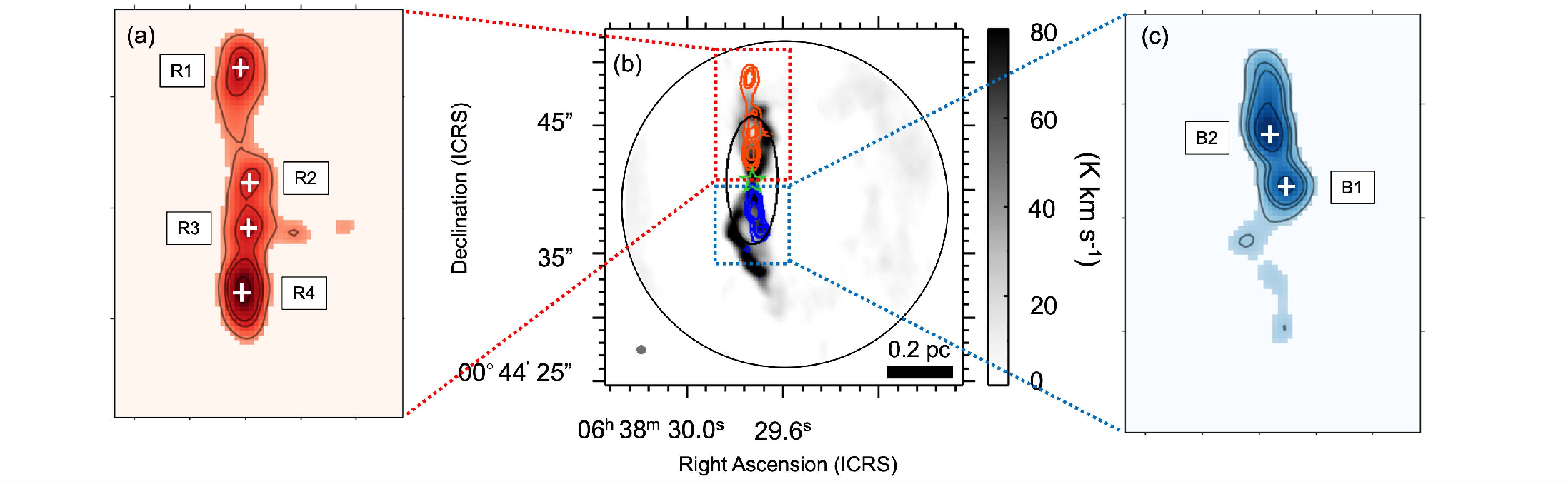}
\caption{(a) The CO($J$ = 3--2) integrated intensity map of the red-shifted jet components is shown. 
The white crosses denote the positions of the jet bullets which are identified from the emission peaks in the integrated intensity map of the jet components.
The number labels represent the emitting order of the jet bullets.
(b) The integrated intensity map of the outflows and jets is presented.
The color map and contour lines represent the outflows and jets, with blue and red contours indicating the blue-shifted and red-shifted components, respectively. 
The contour levels are set at 5, 12, 16, 28, and 50$\sigma$ of the rms noise, where 1$\sigma$ corresponds to 0.86 K km s$^{-1}$ and 1.00 K km s$^{-1}$ for the red and blue lobes, respectively.
The green star indicates the position of the continuum peak.
The synthesized beam size is indicated in the bottom left corner of the map with a gray ellipse.
The black circle represents the field of view.
The black ellipse shows the region from which the spectra are extracted in Figure \ref{image_co_spec}.
(c) The CO($J$ = 3--2) integrated intensity map of the blue-shifted jet components is shown. 
The white crosses denote the positions of the jet bullets which are identified from the emission peaks in the integrated intensity map of the jet components.
The number labels represent the emitting order of the jet bullets.
}
\label{image_co}
\end{center}
\end{figure*}

 \begin{figure*}[tp!]
\begin{center}
\includegraphics[width=16.0cm]{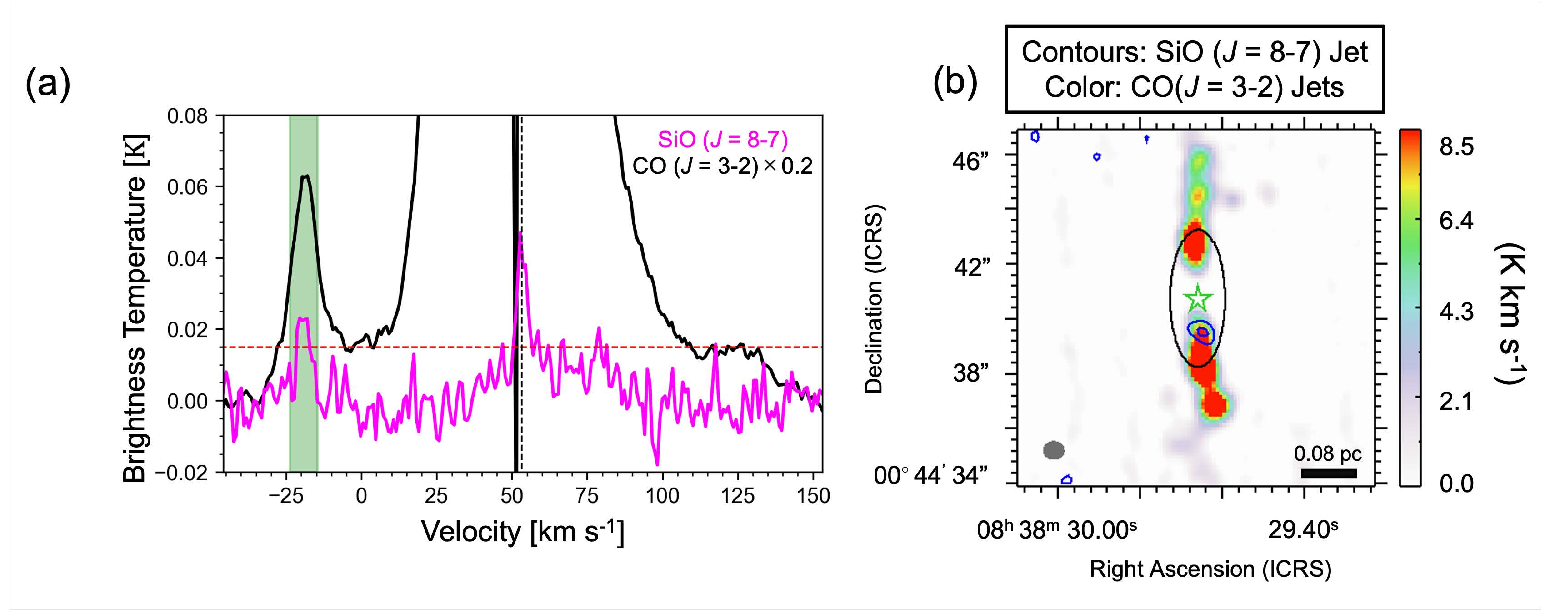}
\caption{(a) The spectra of SiO($J$ = 8--7) (magenta) and CO($J$ = 3--2) (black) are shown.
The spectra are extracted from a $2\farcs0 \times 5\farcs0$ elliptical region centered on the protostar, as indicated by the black ellipse in panel (b).
The green-shaded area represents the jet component.
The red and black dotted lines represent the 3$\sigma$ rms noise level (1$\sigma$ = 0.005 K) of SiO($J$ = 8--7) and the systemic velocity of Sh 2--283--1a SMM1 (53.2 km s$^{-1}$), respectively.
(b) The integrated intensity map of SiO($J$ = 8--7) and jet components of CO($J$ = 3--2) are shown.
The blue contours and color in the intensity map correspond to the jet components of SiO($J$ = 8--7) and CO($J$ = 3--2), respectively. 
The contour levels are at 3 and 6$\sigma$ of the rms noise level, where 1$\sigma$ is 0.24 K km s$^{-1}$.
The black star indicates the emission peak of the 0.87-mm continuum.
The synthesized beam size is depicted in the bottom left corner as a gray ellipse.
}
\label{image_SiO}
\end{center}
\end{figure*}

 \begin{figure*}[tp!]
\begin{center}
\includegraphics[width=16.0cm]{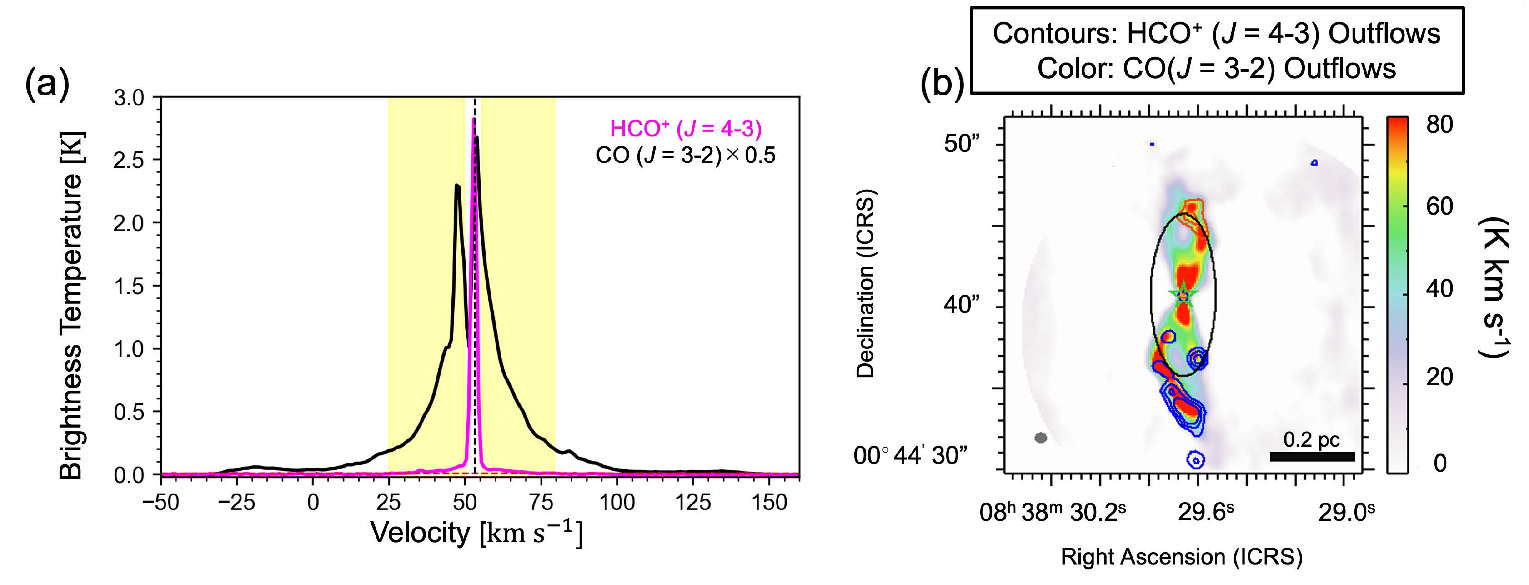}
\caption{(a) The spectra of HCO$^+$($J$ = 4--3) (magenta) and CO($J$ = 3--2) (black) are shown.
The spectra are extracted from a $4\farcs0 \times 10\farcs0$ elliptical region centered on the protostar, as indicated by the black ellipse in panel (b).
In the spectrum, the yellow-shaded area represents the outflow components.
In the spectrum, the red and black dotted lines represent the 3$\sigma$ rms noise level (1$\sigma$ = 0.003 K) of HCO$^+$($J$ = 4--3) and the systemic velosity of Sh 2--283--1a SMM1 (53.2 km s$^{-1}$), respectively.
(b) The integrated intensity map of HCO$^+$($J$ = 4--3) and CO($J$ = 3--2) outflow components are shown.
The contours and color in the intensity map correspond to the outflow components of HCO$^+$($J$ = 4--3) and CO($J$ = 3--2). 
The blue and red contours indicate the blue- and red-shifted components,respectively.
The contour levels are at 3, 6, and 10$\sigma$ of the rms noise level, where 1$\sigma$ is 1.1 K km s$^{-1}$ and 1.0 K km s$^{-1}$ for the blue- and red-shifted components, respectively.
The green star indicates the position of the continuum peak.
The synthesized beam size is depicted in the bottom left corner as a gray ellipse.}
\label{image_HCO}
\end{center}
\end{figure*}


\section{Results and Analysis} \label{sec_res} 
As summarized in Table \ref{ta_source_prop}, outflows were detected in 5 out of the 16 protostellar candidates (Sh 2--283--1a SMM1, NOMF05-16--1, NOMF05-23--1, NOMF05-23--3, and NOMF05-63--1).
In this paper, we mainly focus on Sh 2--283--1a SMM1 since it is the only source which we detect the distinct jet components.
The other 4 outflow sources are discussed in Section \ref{sec_others}.
The basic properties of the outflow driving sources are summarized in Table \ref{ta_source_prop}.
The information of the molecular transitions for detecting the outflows and jets is summarized in Table \ref{tab_linfo}



\begin{deluxetable}{ l c c c c}
\tablecaption{Molecular Transitions \label{tab_linfo}}
\tabletypesize{\footnotesize} 
\tablehead{
\colhead{Molecule}   &\colhead{Transition} &\colhead{Rest Frequency}   & \colhead{$E_u/k$} \\ 
\colhead{}& \colhead{} & \colhead{(GHz)}   &   \colhead{(K)} 
}
\startdata 
CO & $J$ = 3--2& 345.7959899 &  33.2    \\
 SiO& $J$ = 8--7&347.3305810 & 75.0    \\
  HCO$^+$&  $J$ = 4--3 & 356.7342230 & 42.8   \\
\enddata
\end{deluxetable}

\begin{deluxetable}{ l c c c c}
\tablecaption{Observed Properties of the Outflows and Jets in Sh 2--283--1a SMM1
\label{tab_measure}}
\tabletypesize{\footnotesize} 
\tablehead{
\colhead{Molecule}   & \colhead{Velocity Range} & \colhead{Median Velocity}  & \colhead{Extent}  \\ 
\colhead{}& \colhead{(km s$^{-1}$)}   &   \colhead{(km s$^{-1}$)}  &    \colhead{(pc)} 
}
\startdata 
CO($J$ = 3--2) &    &    &   \\
 Outflow--red&[55.7, 103.2]&  26.3&  0.29      \\
  Outflow--blue&[-1.3, 50.1]& 28.8&  0.36      \\
Jet--red &[103.3, 143.2]& 70.1&  0.33       \\
 Jet--blue&[-31.0, -1.4] &69.4&   0.31   \\
\hline
 SiO($J$ = 8--7)& &  &   &  \\
 Jet--blue &[-24.0, -14.0]&72.2& 0.08     \\
 \hline
  HCO$^+$($J$ = 4--3)&   &  &  &  \\
 Outflow--red&[55.7, 80.0]&14.6& 0.24   \\
  Outflow--blue&[25.2, 50.1]& 15.6 & 0.43    \\
\enddata
\tablecomments{(1) -red and -blue mean the red- and blue-shifted components.\\
(2) Median velocity is the relative value to the systemic velocity of the source (53.2 km s$^{-1}$).}
\end{deluxetable}

\subsection{CO outflows and jets} \label{sec_line_co}
We here analyze the multiple velocity components of the outflows and jets observed in Sh 2--283--1a SMM1.  
%
Figure \ref{image_co_spec} shows the CO($J$ = 3--2) spectra extracted from the 4.0$\arcsec$ $\times$ 10.0$\arcsec$ (3.2 $\times$ 10$^4$ au $\times$ 7.9 $\times$ 10$^4$ au at 7.9 kpc from the Sun) elliptical region around the emission peak of the 0.87-mm continuum.
We find that there are two components on both the red- and blue-shifted sides: the two peaks marked in yellow correspond to the outflow emission and the other two peaks marked in green correspond to the jet emission, respectively.
The detailed information about the velocity regimes of the outflows and jets is summarized in Table \ref{tab_measure}.
The velocity criteria for separating the core and outflow components are based on the Full Width at Tenth Maximum (FWTM = 1.8 $\times$ FWHM) of the HCO$^+$($J$ = 4–3) line profile.
The FWHM of the HCO$^+$($J$ = 4–3) line was derived by performing a Gaussian fitting to the HCO$^+$ spectrum extracted from a 0$\farcs$89 × 0$\farcs$70 elliptical region centered at the peak of the 0.87-mm continuum emission.
This small region was chosen to minimize contamination from the outflow wing components.
This is similar to the method used by \citet{Maud2015}, where the FWTM of C$^{18}$O is used to separate the core velocity components from the outflow velocity components.
The velocity components larger than this FWTM are identified as outflow components. 
The terminal velocities of the outflows are determined by the morphology of the emission distribution and the line shape.
For Sh 2--283--1a SMM1, bullet structures are appearing at the line of sight relative velocity of $V_{\rm{obs}}$ $\sim$50 km s$^{-1}$ for both red- and blue-shifted components in the channel maps, which correspond to the emission bottoms of the outflow peaks in Figure \ref{image_co_spec}.
The terminal velocities of the jets are determined whether the line intensity is below 3$\sigma$ of the rms noise level.

Integrated intensity map is obtained by integrating these velocity components for CO($J$ = 3--2) as shown in Figure \ref{image_co} (b). 
This figure clearly shows the presence of both the bipolar outflows and jets which are often detected in nearby protostellar sources \citep[e.g.,][]{bachiller90,Lee2000, Santiago2009, Hirano2010, Tafalla2010, Matsushita2019, Łukasz2019}.
The orientation of the outflows and jets coincides with each other and they are seemed to be within the field of view.
The red-shifted component of the jets extends toward the edge of the field of view, where the sensitivity is $\sim$3 times lower compared to the central region. 
Consequently, some flow components may be missed in the outer regions of the observations.
We find four jet bullets in the red-shifted side and two in the blue-shifted side. 
This is the first time that such jets as well as bullet structures are detected and spatially resolved in a low-metallicity environment. 



\subsection{SiO jet} \label{sec_line_sio} 
In addition to CO, a jet component is also detected on the blue-shifted side with the SiO($J$ = 8--7) line (Figure \ref{image_SiO} (a)).
The velocity peak of this SiO jet component ($V_{\rm{obs}}$ $\sim$70 km s$^{-1}$) well coincides with the CO blue-shifted jet as shown in this figure.
The observed properties of this SiO jet component are summarized in Table \ref{tab_measure}.
The velocity range was derived in a similar manner to that of the CO jet (Section \ref{sec_line_co}).
We extracted the spectra from a 0$\farcs$70-diameter region centered on the emission peak of the SiO jet and fitted a Gaussian profile to the extracted spectra.
The velocity range of the jet component was then defined based on this fitted line profile.

SiO is thought to be produced in shocked regions through the sputtering of Si-bearing species in dust grains and chemical transformation to SiO in the gas phase \citep{Sch97_SiO}.
Thus, SiO is known to be a tracer of shocked gas and often traces jet components which continuously drive shocks along their paths \citep[e.g.,][]{Sch97_SiO,Nisini2007,Tafalla2010,Takahashi2024}.
The coincidence in both the spatial distribution and relative velocity of the SiO jet and the CO jet components would indicate a physical association between these two components.


\subsection{HCO$^+$ Outflow} \label{sec_line_HCO} 
Outflow components are also detected with HCO$^+$($J$ = 4--3) on both the red- and blue-shifted lobes (Figure \ref{image_HCO}).
The HCO$^+$ outflows have lower terminal velocities compared to the CO outflows (see Figure \ref{image_HCO} (a)).
The distribution of these outflow components shows fragmentary structures and they seem to trace the edge of the CO outflows (see Figure \ref{image_HCO} (b)).
Since the critical density ($n_{\rm{cr}}$) and upper state energy ($E_{\rm u}$) of HCO$^+$($J$ = 4--3) ($n_{\rm{cr}}$ $\sim$10$^5$ cm$^{-3}$ and $E_{\rm u}$ = 42.8 K) are relatively high compared to those of CO($J$ = 3--2) ($n_{\rm{cr}}$ $\sim$10$^3$ cm$^{-3}$ and $E_{\rm u}$ = 33.2 K), these HCO$^+$($J$ = 4--3) outflow components would trace the high density region of the CO($J$ = 3--2) outflows.

\subsection{Various molecular lines in Sh 2--283--1a SMM1} \label{sec_chem}
A variety of molecular species including complex organic molecules (COMs; organic molecules which consist of at least 6 atoms) are detected from the emission peak of 0.87-mm continuum (see Figure \ref{image_spec_sum}). 
This includes multiple high--excitation lines ($E_u$ $\geq$ 100 K) of CH$_3$OH and SO$_2$.
It indicates that warm molecular gas is surrounding the protostar at the position of Sh 2--283--1a SMM1.
Such a chemically-rich region is known as a hot core, which is a key signature of massive star formation activity.
This is the second detection of a hot core in the outer Galaxy after WB89--789 SMM1 reported in \citet{Shimonishi2023}. 
The detailed analyses of the chemical composition of the Sh 2--283--1a SMM1 hot core will be presented in a forthcoming paper (Ikeda et al. in prep.). 



\begin{deluxetable*}{ l c c c |c c c c c}
\tablecaption{Dynamical Properties of Outflows and Jets in Sh 2--283--1a SMM1 \label{ta_flow_dy} }
\tabletypesize{\footnotesize} 
\tablehead{ &  &  \colhead{Outflows} &  & & \colhead{Jets} \\ \colhead{Parameters}   &  \colhead{Red}                  & \colhead{Blue}  & \colhead{Total}  &  \colhead{Red}                  & \colhead{Blue}  & \colhead{Total}
}
\startdata 
Mass (M$_\sun$) &  31 & 42 & 73 &    1.9  & 1.9 & 3.8\\
Median Velocity (km s$^{-1}$)$^a$ & 37.2 & 40.7  & --   & 99.1 & 98.1 & -- \\
Momentum (M$_\sun$ km s$^-1$)$^a$ &1.1 $\times$ 10$^{3}$  &1.7 $\times$ 10$^{3}$ &  2.8 $\times$ 10$^{3}$& 1.8 $\times$ 10$^{2}$  & 1.8 $\times$ 10$^{2}$  &  3.6 $\times$ 10$^{2}$  \\
Dynamical Timescale (yr)$^a$ &  1.1 $\times$ 10$^{4}$ & 1.2 $\times$ 10$^{4}$ & -- &  4.6 $\times$ 10$^{3}$ & 4.4 $\times$ 10$^{3}$ & -- \\
 Mass Loss Rate (M$_\sun$ yr$^{-1}$)$^a$    & 2.8 $\times$ 10$^{-3}$  & 3.4 $\times$ 10$^{-3}$  &  6.2 $\times$ 10$^{-3}$ & 4.0 $\times$ 10$^{-4}$  & 4.2 $\times$ 10$^{-4}$  &  8.2 $\times$ 10$^{-4}$ \\
 Force (M$_\sun$ km s$^{-1}$ yr$^{-1}$)$^a$ &  0.11 & 0.14  & 0.25  &  4.0 $\times$ 10$^{-2}$ & 4.1 $\times$ 10$^{-2}$  &  8.1 $\times$ 10$^{-2}$ \\
    Energy (erg)$^a$ & 4.2 $\times$ 10$^{47}$ &  6.9 $\times$ 10$^{47}$  &  1.1 $\times$ 10$^{48}$ & 1.8 $\times$ 10$^{47}$ &  1.8 $\times$ 10$^{47}$  &  3.6 $\times$ 10$^{47}$ \\
 Kinetic Luminosity (erg yr$^{-1}$)$^a$  & 3.9 $\times$ 10$^{43}$ &  5.7 $\times$ 10$^{43}$  &  9.6 $\times$ 10$^{43}$ & 3.9 $\times$ 10$^{43}$ &  4.0 $\times$ 10$^{43}$  &  7.9 $\times$ 10$^{43}$ \\
\enddata
\tablecomments{$^a$$i$ = 45$^\circ$ is assumed.
}

\end{deluxetable*}


\section{Discussion} \label{sec_dis} 

\subsection{Comparison of the morphological and dynamical properties of Sh 2--283--1a SMM1 with the nearby protostellar sources} \label{sec_str} 
We compare the dynamical properties of the outflows and jets with CO($J$ = 3--2) in Sh 2--283--1a SMM1 with the protostellar sources located in the solar neighborhood.
Since the outflows and jets are extended broadly in the field of view (see Figure \ref{image_co}), there must be some missing flux of the flow gas spread beyond the Maximum recoverable scale of the observations, resulting in underestimate the flow properties.
In addition, the sensitivity at the edge of the field of view is $\sim$3 times lower than at the center, which could lead to the possibility of missing faint or diffuse components near the edge to be missed, further contributing to the underestimation.
Hence, the dynamical properties of the outflows and jets related to their flow mass (e.g., mass, momentum, energy) discussed below can be considered as lower limits.

First, we calculate the dynamical timescales ($t_{\rm{d}}$) for both the outflows and jets.
$t_{\rm{d}}$ can be derived using the projected outflow length and velocity as follows:
\begin{equation}
t_{\rm{d}} = \left(\frac{L_{\rm{obs}}}{V_{\rm{obs}}}\right)\left(\frac{\cos(i)}{\sin(i)}\right),
\label{Eq_dyn}
\end{equation}
where $V_{\rm{obs}}$ and $L_{\rm{obs}}$ are line of sight gas velocities and the projected length of the outflows and jets, respectively.
$i$ is the inclination angle with respect to the line of sight and we here assume 45$^{\circ}$. 
Although the inclination angle of Sh 2--283--1a SMM1 is unknown, the red- and blue-shifted lobes of both the outflows and jets have clear bipolar structures (Figure \ref{image_co}), which indicate that they are closer to the edge-on geometry rather than the face-on (i.e. $i \geq 40^\circ$).
Note that the inclination angle affects the dynamical properties other than outflow mass.
When $i$ $\geq$ 84$^\circ$, uncertainties in derived dynamical values due to the inclination angle become significant by about one order of magnitude compared to the accurate values.

The dynamical timescales of the outflows are approximately 1.2 $\times$ 10$^4$ years for the red lobe and 1.4 $\times$ 10$^4$ years for the blue lobe, with the assumption of the median flow velocity of each velocity component as the representative flow velocity of the CO($J$ = 3--2) outflows ($V_{\rm{obs}}$ = 26.3 and 28.8 km s$^{-1}$, respectively).
Under the same assumptions, the dynamical timescale of the jets is approximately 5.0 $\times$ 10$^3$ years for both the red and blue lobes, with assuming the representative flow velocities of $V_{\rm{obs}}$ = 70.1 and 69.4 km s$^{-1}$, respectively.
These values are summarized in Table \ref{tab_measure}.

With assuming that CO emission in both the outflow and jet components are optically thin and under the condition of the local thermodynamic equilibrium (LTE), the outflow mass ($M_{\rm{out}}$) is derived by using the following equation:
\begin{equation}
M_{\rm out} =  2 \mu_p m_{\rm{p}} X_{\rm{CO}} \int T_{\rm{CO}}  dV,
\label{Eq_mass}
\end{equation}
where $\mu_p$ is mean atomic mass per H atom (1.41), $m_{\rm{p}}$ is mass of a hydrogen atom, $\int T_{\rm{CO}} dV$ is sum of integrated intensity of CO outflow, and $X_ {\rm{CO}}$ is a conversion factor of CO intensity to H$_2$ column density.
Here, we use a metallicity-corrected $X_\mathrm{CO}$ of 5 $\times$ 10$^{20}$ cm$^{-2}$ (K km s$^{-1}$) from \cite{Bol13}. 
For both the outflow and jet components, we derive the dynamical properties (i.e., outflow momentum, mass loss rate, outflow force, outflow energy, and outflow kinetic luminosity) with assuming $i$ = 45$^\circ$.
The derived dynamical properties for each lobe are summarized in Table \ref{ta_flow_dy}.
These dynamical properties are similar to those of massive protostellar sources detected in the solar neighborhood \citep{Beuther2002,Wu2004,Maud2015}. 

Note that CO($J$ = 3--2) with the LTE assumption is known to potentially underestimate the outflow mass, as this transition is often sub-thermally excited in outflows \citep{Ginsburg2011, Dunham2014}.
\cite{Ginsburg2011} conducted a non-LTE analysis using RADEX \citep{vdT07} and showed that CO($J$ = 3--2) with the LTE assumption can underestimate the outflow mass by up to 1--2 orders of magnitude compared to CO($J$ = 1--0), whose critical density is 27 times lower than that of CO($J$ = 3--2), with the gas densities of 10$^{2-4}$ cm$^{-3}$.
Thus, the derived dynamical properties related to the flow mass would be underestimated by the order of 1--2.
To obtain more accurate estimates of the flow properties for Sh 2–283–1a SMM1 and other newly discovered outflow sources (see Section \ref{sec_others}), follow-up observations using lower-excitation CO lines (i.e., $J$ = 2–1 or $J$ = 1–0) will be necessary.

Although the full spectral energy distribution (SED) of Sh 2--283--1a SMM1 is not available, we estimate its luminosity by scaling the luminosity of the LMC hot core (ST16) with its $K_s$-band magnitude ([$K_s$]) as conducted in \cite{Shimonishi2021}.
The luminosity of ST16 is determined by an infrared dataset from 1 to 1200 $\mu$m. 
The total luminosity of ST16 is 3.1 $\times$ 10$^5$ L$_\sun$ and [$K_s$] is 13.4 mag at 50 kpc.
For Sh 2--283--1a SMM1, [$K_s$] is 15.6 mag at 7.9 kpc.
With scaling the luminosity of ST16 with the [$K_s$] and distance, we obtain 6.7$\times$ 10$^3$ L$_\sun$ as the luminosity of Sh 2--283--1a SMM1.
This value is consistent with the upper limit derived from far-infrared data obtained from the \textit{AKARI} FIS all-sky survey \citep{Yam10}.
Note that this is an estimation from only one infrared data and the obtained value would have large uncertainty.
Future high spatial resolution and multiwavelength observations are highly required.
Moreover, we can roughly estimate the luminosity of Sh 2--283--1a SMM1 by using correlation between outflow mass and luminosity.
\cite{Maud2015} performed a statistical study of protostellar outflows and they found correlations between outflow masses and luminosities (see their Figure 5). 
Using this correlation, the luminosity of Sh 2--283--1a SMM1 is estimated to be about 10$^3$ to 10$^5$ L$_{\sun}$.
The estimated luminosity by two different methods is consistent with each other and Sh 2--283--1a SMM1 would be an intermediate- to high-mass protostellar source.



All these morphological and dynamical properties of Sh 2--283--1a SMM1 resemble those observed in nearby protostellar sources.
It indicates that the driving mechanisms of outflows and jets in a low-metallicity protostellar core in the outer Galaxy are similar to those of the inner Galaxy counterparts. 


\begin{figure*}[tp!]
\begin{center}
\includegraphics[width=18.0cm]{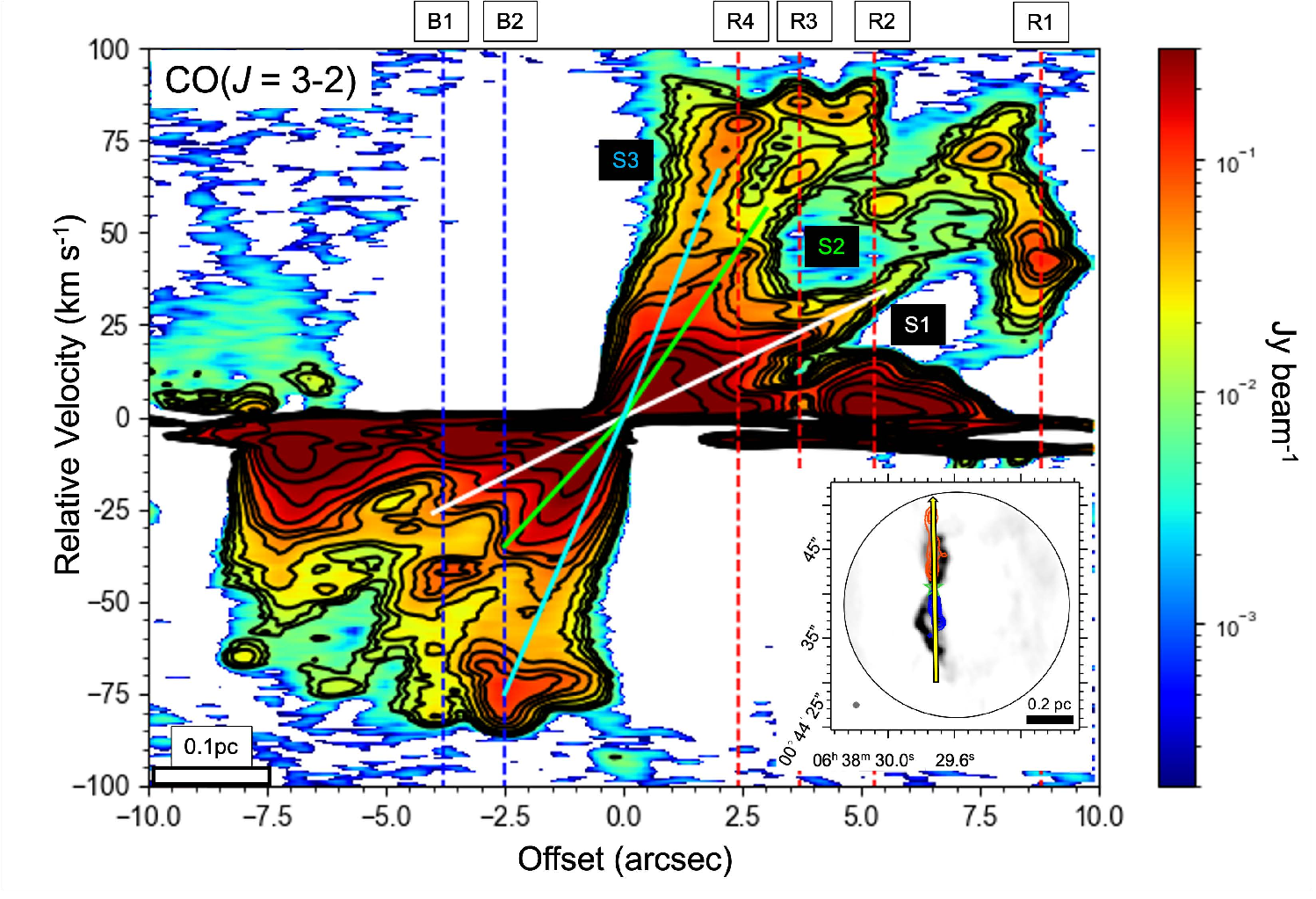}
\caption{The position-velocity diagram of Sh 2--283--1a SMM1 along the flow axis (p.a = 1.1$^\circ$).
The contour levels are at 5, 7, 10, 15, 25, 35, 45, 75, 115, 165, and  300$\sigma$ of the rms noise level, with 1$\sigma$ equals to 1.8 mJy beam$^{-1}$.
The vertical dotted lines to the horizontal axis show the positions of the jet bullets (R1, R2, R3, R4, B1, and B2) in Figure \ref{image_co}.
The solid lines trace the spine-like structures (S1, S2, and S3) discussed in Section \ref{sec_pv_1}.
The right-bottom panel shows the integrated intensity map of the outflows and jets as in Figure \ref{image_co}.
The yellow allow indicates the direction for the slice of the PV diagram.}
\label{im_pv}
\end{center}
\end{figure*}

\subsection{The episodic mass-ejection events in Sh 2--283--1a SMM1} \label{sec_pv_1} 
Figure \ref{im_pv} shows the position--velocity diagram (PV diagram) of Sh 2--283--1a SMM1 along the flow axis (p.a = 1.1$^\circ$).
The vertical dotted lines to the horizontal axis show the positions of the jet bullets (R1, R2, R3, R4, B1, and B2) in Figure \ref{image_co}.
The PV diagram clearly shows that the velocity increases approximately linearly with respect to the distance from the continuum peak.
This morphological characteristic in the PV diagram is commonly observed in nearby outflow sources and is often called as Hubble-like velocity structure or Hubble flow \citep[e.g.,][]{Lee2000,Hirano2010, Nony2020, Takahashi2024}.

Moreover, we find several spine-like structures, which are emphasized by the solid lines (S1, S2, and S3) in Figure \ref{im_pv}.
S2 is connected to R3 and B1, as well as S3 is connected to R4 and B2, while S1 is not connected to any jet bullets.
These spine-like structures associated with the jet bullets would trace their entrained gas or multiple velocity components of the jets.
Such distinctive structures in the PV diagram have also been confirmed in some outflow sources in nearby star-forming regions \citep{Plunkett2015,Nony2020, Takahashi2024}.

\cite{Takahashi2024} observed an extremely young intermediate-mass protostellar source, MMS 1/OMC-3, and detected both the outflows and jets with CO($J$ = 2--1) and SiO($J$ = 5--4), respectively.
Their PV diagram cut along the outflow axis (see their Figure 5) shows similar structures that are presented in Figure \ref{im_pv}.
By comparing with numerical calculations made by \cite{Machida2019} and presented in \cite{Takahashi2024}, the spine-like structures indicate the episodic mass-ejection which is related to each mass-accretion event to the central protostar. 

Although the spatial resolution of this work ($\sim$5000 au) is approximately two orders of magnitude smaller than that of \citet{Takahashi2024} ($\sim$80 au), the velocity structures detected in the PV diagram are remarkably similar, suggesting that both studies trace the same phenomenon related to episodic mass-ejection events.
This similarity may be attributed to the difference in dynamical timescales between the two sources. 
In \citet{Takahashi2024}, the jets have a dynamical timescale of $\sim$50 years, which is significantly shorter than that of Sh 2--283--1a SMM1 ($t_{\rm{d}}$$\sim$4000 years). 
Therefore, it is plausible that such jet structures evolve over time and can remain detectable even on larger spatial scales.
Furthermore, similar spine-like structures have also been identified at comparable spatial resolution ($\sim$5000 au) in nearby star-forming regions \citep{Nony2020}, supporting the interpretation that episodic ejection features are observable even with the spatial resolution available in this study. 

Since the inverse of the slope of the spine-like structures corresponds to their time scales, we can estimate the time intervals of the episodic mass-ejection events (jet episodicity) with the slope of each spine-like structure in Figure \ref{im_pv}.
The slopes of the red-shifted side of S1, S2, and S3 are 6.2, 18.8, and 33.3 km s$^{-1}$ arcsec$^{-1}$, and they correspond to the time scales of approximately 6000, 2000, and 1100 years.
For the blue-shifted side of S1, S2, and S3, the slopes are 6.5, 14.0, and 30.0 km s$^{-1}$ arcsec$^{-1}$ and they correspond to the time scales of approximately 5800, 2700, and 1200 years.
Thus, the time intervals of the episodic mass-ejection events for Sh 2--283--1a SMM1 are predicted to be on average 1600 years (900--4000 years).
This value is comparable to those of the inner Galactic sources \citep[e.g., 290--4300 years with 72 bullets, with a median of $\sim$790 years;][]{Nony2020}.

\subsection{SiO abundance in the jet component}
\label{sec_sio_abun} 
\subsubsection{Derivation of molecular column densities} \label{sec_derive} 
Since we detect the jet components with CO and SiO lines, we can estimate the column densities of these molecular species and compare the chemical properties of the jet components with nearby jet sources.
Here, we extract the spectra from a circular region (a radius of 0$\farcs$35) at the emission peak of the SiO jet component (see Figure \ref{image_co}).
For simplicity, we assume that both the CO and SiO emission in the jet components are optically thin and under the LTE condition.
We estimate the column densities by using the following formula \citep[e, g.,][]{Mangum2015}:
\begin{equation}
N_{u} = \frac{3 k \int T_{\mathrm{b}} \, dV \, g_{u} \, Q({T_{rot})}}{8 \pi^{3} \nu S \mu^{2}} \exp\left(\frac{E_{u}}{k T_{rot}}\right),
\label{Eq_col}
\end{equation}
where $N_u$ is the column density of the molecules in the upper state, $k$ is the Boltzmann constant, $T_\mathrm{b}$ is the brightness temperature, g$_u$ is the degeneracy of the upper state, $T_{\rm{rot}}$ is the rotational temperature, Q($T_{\rm{rot}}$) is the partition function at $T_{\rm{rot}}$, $\nu$ is the rest frequency, $S$ is the line strength, $\mu$ is the dipole moment, and $E_{\rm u}$ is the upper state energy.
All the spectroscopic parameters (Q($T_{\rm{rot}}$), g$_u$, $\nu$, $\mu$, $E_{\rm u}$, and $S$) are extracted from the Cologne Database for Molecular Spectroscopy (CDMS) \citep{Muller2001, Muller2005}.
Since we detect only one transition for each molecule in this work, the rotational temperatures ($T_{\rm{rot}}$) of CO and SiO are unknown. 
We therefore estimate $T_{\rm{rot}}$ based on the literature and compare the $N\mathrm{(SiO)}$/$N\mathrm{(CO)}$ ratio of the jet component with that of nearby outflow sources.

\subsubsection{Comparison of SiO abundance with nearby sources} \label{sec_abun_comp} 
%
Since we found no high-mass counterparts ($>$10$^3$ L$_{\sun}$) for comparing the $N\mathrm{(SiO)}$/$N\mathrm{(CO)}$ ratio in the jet components.
Consequently, the $N\mathrm{(SiO)}$/$N\mathrm{(CO)}$ ratio is compared with those of low-mass sources.

A nearby low-mass protostellar source, L1448-mm (L1448), is known to have jet bullet structures \citep{bachiller90}.
\citet{Nisini2007} derived the rotational temperature of the SiO jet components with the multiwavelength observations. 
The SiO multi-line analysis was conducted (from $J = 2-1$ to $J = 11-10$), and the derived rotational temperatures and column densities are $T_{\rm{rot}}$(SiO) = 100--500 K and $N\mathrm{(SiO)}$ = 3 $\times$ 10$^{13}$ cm$^{-2}$, respectively.
\citet{Tafalla2010} derived $T_{\rm{rot}}$(CO) = 11 K and $N\mathrm{(CO)}$ = 5.7 $\times$ 10$^{15}$ cm$^{-2}$ for this jet source (from $J$ = 1--0 and 2--1).
From these results, the $N\mathrm{(SiO)}$/$N\mathrm{(CO)}$ of L1448 is 5 $\times$ 10$^{-3}$.
Both \cite{Nisini2007} and \cite{Tafalla2010} conducted the observations using single-dish telescopes (the IRAM 30-m telescope and the James Clerk Maxwell Telescope) with a spatial resolution of $\sim$5000 au, which is comparable to this work.
For a fair comparison, we assume that the rotational temperatures of the jet components of SiO and CO are 100--500 K and 11 K, respectively.
Under these assumptions, we derive $N\mathrm{(SiO)}$/$N\mathrm{(CO)}$ = (1.1--3.2) $\times$ 10$^{-4}$, 
where $N\mathrm{(SiO)}$ = (1.9--5.2) $\times$ 10$^{12}$ cm$^{-2}$ and $N\mathrm{(CO)}$ = 2.1 $\times$ 10$^{16}$ cm$^{-2}$, respectively. 
This $N\mathrm{(SiO)}$/$N\mathrm{(CO)}$ ratio is more than an order of magnitude lower than that of L1448. 

\cite{Dutta2024} found 17 SiO jet driving sources in 39 protostars in the Orion molecular cloud.
They derived the column densities of CO and SiO under the assumption of LTE, optically thin emission, and $T_{\rm rot}$ of 150 K for both CO and SiO.
The derived $N\mathrm{(SiO)}$/$N\mathrm{(CO)}$ ranges 5.3--88.4 $\times$ 10$^{-4}$ and the averaged $N\mathrm{(SiO)}$/$N\mathrm{(CO)}$ is 25.5 $\times$ 10$^{-4}$. 
For Sh 2--283--1a SMM1, applying the same assumption (i.e., LTE condition, optically thin emission, and $T_{\rm rot}$ of 150 K for both CO and SiO), the derived $N\mathrm{(SiO)}$/$N\mathrm{(CO)}$ is 1.7 ($\pm$ 0.6) $\times$ 10$^{-4}$.
This result is close to the minimum value of \cite{Dutta2024} and about an order of magnitude lower than their average value.

The present analysis suggests the low gas-phase SiO/CO ratio in the jet bullet of Sh 2--283--1a SMM1, but the reason for this remains unclear. 
It is believed that energetic processes in the shock regions such as the sputtering or photolysis of dust grains help release silicon or silicon-bearing molecules into the gas-phase, leading to the efficient formation of SiO \citep[e.g.,][]{Sch97_SiO, Tab20}. 
Since the kinematic and morphological properties of the outflows and jets in Sh 2--283--1a SMM1 are similar to those of nearby sources, we speculate that the different shock chemistry may be related to the different dust composition in a unique low-metallicity environment of the outer Galaxy. 

The luminosity difference between Sh 2--283--1a SMM1 and its counterparts may also contribute to the observed variation in the $N\mathrm{(SiO)}$/$N\mathrm{(CO)}$ ratio. 
As mentioned above, we cannot find the nearby high-mass protostellar sources for which both CO and SiO abundances in the jet components are measured. 
Since it is known that the dynamical properties of outflows have strong correlation to their luminosity \citep[e.g.,][]{Wu04, Maud2015}, luminosity may affect the shock chemistry induced by outflows and jets.
The luminosity of the counterparts to which we compared the $N\mathrm{(SiO)}$/$N\mathrm{(CO)}$ ratio above is on the order of 10$^{-1}$ to 10$^{2}$ L$_{\sun}$, which is more than one order of magnitude lower than that of Sh 2--283--1a SMM1 (L $\sim$ 6.6 $\times$ 10$^3$ L$_{\sun}$, see Section \ref{sec_str}).
We need to increase the number of high-mass outflow/jet sources to better understand the luminosity effect on the shock chemistry.

Although the $N\mathrm{(SiO)}$/$N\mathrm{(CO)}$ ratio derived under the LTE assumption is lower than those reported for nearby sources, non-LTE calculations using RADEX yield different results.
We performed non-LTE analyses for SiO($J$ = 8--7) and CO($J$ = 3--2), assuming kinetic temperatures ($T_\mathrm{kin}$) of 100--500 K and 11 K, respectively.
For other input parameters, we adopted an H$_2$ density ($n(\mathrm{H_2})$) of 10$^{5}$--10$^{6}$ cm$^{-3}$, representing the typical gas density of jet bullets \citep[e.g.,][]{Nisini2007, Lefloch2015}.
The linewidths and peak intensities of SiO and CO were set to 8.5 and 9.1 km s$^{-1}$, and 0.15 and 1.45 K, respectively, as derived from the 0$\farcs$70-diameter aperture analysis described in Section \ref{sec_derive}.
The background temperature was set to 2.73 K.

The resultant $N\mathrm{(SiO)}$/$N\mathrm{(CO)}$ ratio is 4.8--48$\times$10$^{-4}$ with $n(\mathrm{H_2})$ = 10$^{5}$ cm$^{-3}$, while it is 0.9--3.4$\times$10$^{-4}$ with $n(\mathrm{H_2})$ = 10$^{6}$ cm$^{-3}$.
Thus, the derived $N\mathrm{(SiO)}$/$N\mathrm{(CO)}$ ratio strongly depends on the assumed $n(\mathrm{H_2})$.
When $n(\mathrm{H_2})$ = 10$^{5}$ cm$^{-3}$, the $N\mathrm{(SiO)}$/$N\mathrm{(CO)}$ ratio is comparable to that of the nearby counterpart (L1448), whereas with $n(\mathrm{H_2})$ = 10$^{6}$ cm$^{-3}$, the value becomes similar to that of Sh 2--283--1a SMM1 under the LTE assumption.
We also performed non-LTE calculations assuming a uniform $T_\mathrm{kin}$ = 150 K for both CO and SiO.
In this case, the resultant $N\mathrm{(SiO)}$/$N\mathrm{(CO)}$ ratio is 6.6$\times$10$^{-3}$ for $n(\mathrm{H_2})$ = 10$^{5}$ cm$^{-3}$, and 3.3$\times$10$^{-4}$ for $n(\mathrm{H_2})$ = 10$^{6}$ cm$^{-3}$.

These results imply that non-LTE effects should be carefully considered when interpreting molecular abundances in protostellar jets, as the $N\mathrm{(SiO)}$/$N\mathrm{(CO)}$ ratio strongly depends on $n(\mathrm{H_2})$.
Thus, caution must be exercised when directly comparing $N\mathrm{(SiO)}$/$N\mathrm{(CO)}$ ratio derived under LTE assumptions among different sources without considering their excitation conditions.
Multi-line observations of SiO and CO are necessary to understand the excitation conditions of these molecules in the jets.
%

\subsection{The other outflow sources in the outer Galaxy} \label{sec_others} 
\begin{figure*}[tp!]
\begin{center}
\includegraphics[width=18.0cm]{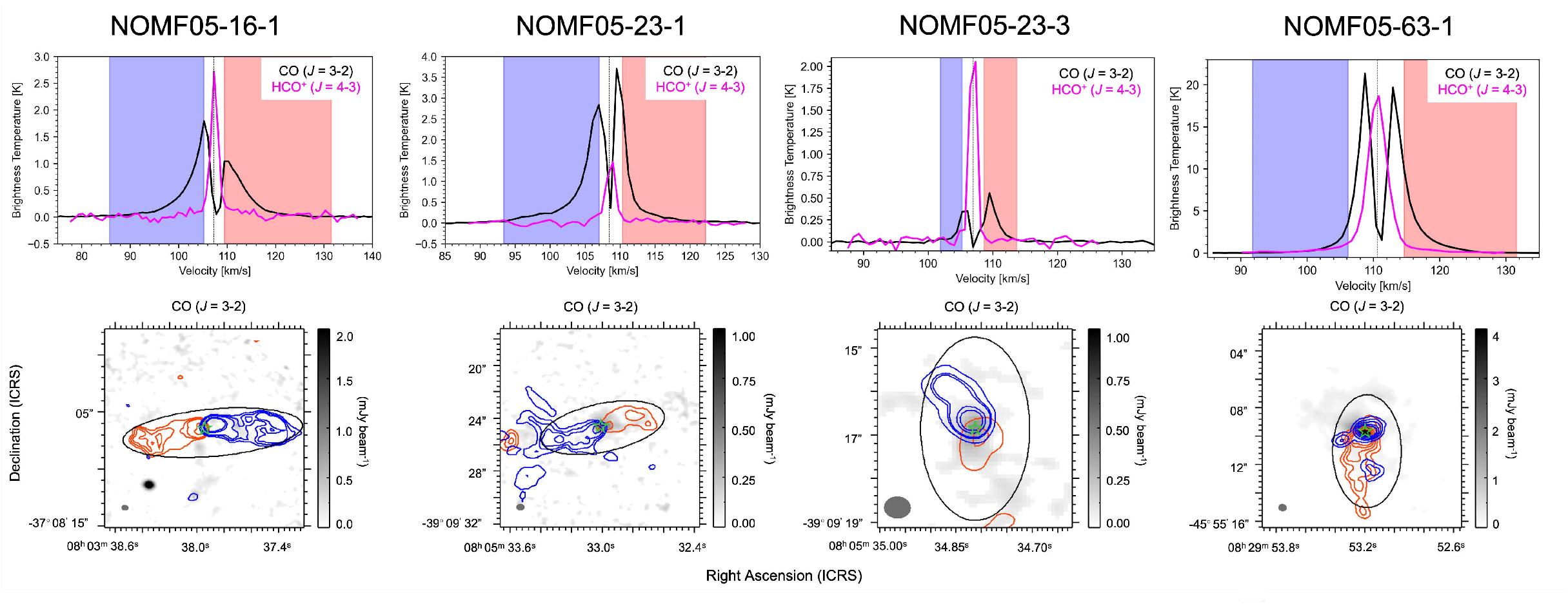}
\caption{Newly detected outflow sources in the outer Galaxy.
The upper panels represent the spectra of CO($J$ = 3--2) (black) and HCO$^+$($J$ = 4--3) (magenta).
The spectra of CO($J$ = 3--2) are extracted from the elliptical region as shown in the lower panels, while the spectra of HCO$^+$($J$ = 4--3) are extracted from the compact region ($\sim$beam size) of the core center for each source.
The blue and red shaded areas in the spectra indicate the velocity ranges used for the integrated intensity maps in the lower panels, corresponding to the blue- and red-shifted lobes, respectively.
The black dotted lines in each spectrum represent the systemic velocities of each source.
The contour levels for NOMF05-16–1 are at 3, 6, 9, 13, and 15$\sigma$ of the rms noise level, with 1$\sigma$ equal to 2.90 K km s$^{-1}$ and 2.07 K km s$^{-1}$ for the red and blue lobes, respectively.
For NOMF05-23–1, the contour levels are at 3, 6, 10, 18, 25$\sigma$ of the rms noise, with 1$\sigma$ equal to 2.00 K km s$^{-1}$ and 0.50 K km s$^{-1}$ for the red and blue lobes, respectively.
The contour levels for NOMF05-23–3 are at 3, 6, 7, 12, and 15$\sigma$ of the rms noise, where 1$\sigma$ is 0.48 K km s$^{-1}$ and 0.35 K km s$^{-1}$ for the red and blue lobes, respectively.
For NOMF05-63–1, the contour levels are at 3, 8, 15, 25, and 35$\sigma$ of the rms noise, with 1$\sigma$ equal to 1.42 K km s$^{-1}$ and 0.86 K km s$^{-1}$ for the red and blue lobes, respectively.
In each integrated intensity map, the gray ellipse indicates the beam size of the observations, and the color maps represent the continuum emission.
The green star in each integrated intensity map represents the  emission peak of the 0.87-mm continuum except for NOMF05-16-1.
For NOMF05-16-1, because of no detection of the continuum emission, the green star represents the emission peak of HCO$^+$($J$ = 4--3).
}
\label{image_pv}
\end{center}
\end{figure*}

We also detected 4 outflow sources besides Sh 2--283--1a SMM1 (Figure \ref{image_pv}).
The plotted CO($J$ = 3--2) spectra in the figure are extracted from the elliptical regions that cover the outflow emitting area of each source.
The CO($J$ = 3--2) integrated intensity maps of NOMF05-16--1, 23--1, and 23--3 clearly show the bipolar structure for both the red- and blue-shifted components, while NOMF05-63--1 shows no clear bipolar structure. 
Although the red-shifted CO($J$ = 3--2) component of NOMF05-63--1 shows the extended structure, its blue-shifted component has the emission peak at the continuum center.
Such monopolar outflows are often detected in the solar neighborhood, and they are believed to arise from various factors, such as obscuration by dense ambient material, asymmetric environments, deflection due to interaction with surrounding gas, or time-variable outflow activity \citep[e.g.,][]{Wu04, Fern2013, Louvet2018}.
All but NOMF05-16--1 are associated with 0.87-mm continuum point sources. 

For these outflow sources, we derive the dynamical properties using the same way as in Section \ref{sec_str}, assuming $i$ = 45$^\circ$.
The derived values are summarized in Table \ref{ta_othrflow_dy}.
The median velocities of these outflows ($\sim$5 to 19 km s$^{-1}$) are relatively low compared to those of Sh 2--283--1a SMM1 ($\sim$37 to 40 km s$^{-1}$).
The outflow masses vary depending on the source, and all of them are lower compared to Sh 2--283--1a SMM1.
The dynamical timescales of these outflows are $\sim$10$^4$ years, consistent with those of 2--283--1a SMM1.

We do not detect any jet component and SiO emission in these 4 sources.
This would be due to the insufficient sensitivity to detect the relatively low intensity emission of their jet components compared with the outflow components (see Figure \ref{image_co_spec}).
Future high sensitivity and spatial resolution observations are strongly needed to investigate their jet structures or protoplanetary disks in the outer Galaxy.


\begin{deluxetable*}{l c c c c c c c c c}
\tablecaption{Dynamical Properties of Outflows in the other four the outer Galaxy sources}
\tabletypesize{\footnotesize} 
\tablehead{ \colhead{Source Name} &  \colhead{Mass} & \colhead{Median}& \colhead{Momentum$^a$} &\colhead{Dynamical} & \colhead{Mass Loss} & \colhead{Force$^a$}   &  \colhead{Energy$^a$}   & \colhead{Kinetic} \\
\colhead{} &\colhead{} &\colhead{Velocity$^a$} & \colhead{} & \colhead{Timescale$^a$} &
\colhead{Rate$^a$} & \colhead{} &\colhead{} & \colhead{Luminosity$^a$} &
\\  \colhead{} & \colhead{(M$_\sun$)} & \colhead{(km s$^{-1}$)}
& \colhead{(M$_\sun$ km s$^{-1}$)} & \colhead{(yr)} & \colhead{(M$_\sun$ yr$^{-1}$)} & \colhead{(M$_\sun$ km s$^{-1}$ yr$^{-1}$)} & \colhead{(erg)} & \colhead{(erg yr$^{-1}$)} 
}
\startdata 
NOMF05-16--1 \\  Red &20 &18.7 &3.7 $\times$ 10$^{2}$& 3.9 $\times$ 10$^{4}$ &5.2 $\times$ 10$^{-4}$& 9.6 $\times$ 10$^{-3}$& 6.9 $\times$ 10$^{46}$ & 1.8 $\times$ 10$^{42}$&   \\ Blue& 23 & 17.0 & 3.8 $\times$ 10$^{2}$& 5.2 $\times$ 10$^{4}$ &4.4 $\times$ 10$^{-4}$& 7.4$\times$ 10$^{-3}$& 6.5 $\times$ 10$^{46}$ & 1.2 $\times$ 10$^{42}$ &\\ Total &45 & -- & 7.6 $\times$ 10$^{2}$&-- &9.6 $\times$ 10$^{-4}$ & 1.7 $\times$ 10$^{-2}$& 1.3 $\times$ 10$^{47}$ &3.0 $\times$ 10$^{42}$ &\\
\hline
NOMF05-23--1 \\  Red & 2.6 & 9.1& 2.4 $\times$ 10$^{1}$ &5.6 $\times$ 10$^{4}$ &4.7 $\times$ 10$^{-5}$ & 4.2 $\times$ 10$^{-4}$ & 2.1 $\times$ 10$^{45}$& 3.8 $\times$ 10$^{40}$ &          
\\ Blue & 12 & 9.6 & 1.1 $\times$ 10$^{2}$& 3.0 $\times$ 10$^{4}$ & 3.9 $\times$ 10$^{-4}$ & 3.8 $\times$ 10$^{-3}$ & 1.1 $\times$ 10$^{46}$  & 3.6 $\times$ 10$^{41}$ &\\ Total &15 &-- &1.3 $\times$ 10$^{2}$ & --&4.4 $\times$ 10$^{-4}$ &4.2 $\times$ 10$^{-3}$& 1.3 $\times$ 10$^{46}$& 4.0 $\times$ 10$^{41}$ &\\
\hline
NOMF05-23--3 \\  Red & 0.11  & 5.8 & 0.63 & 1.9 $\times$ 10$^{4}$& 5.8 $\times$ 10$^{-6}$ &3.4 $\times$ 10$^{-5}$&3.6 $\times$ 10$^{43}$ &  1.9 $\times$ 10$^{39}$ &  \\ Blue & 0.14 &  4.7  & 0.64 & 3.4 $\times$ 10$^{4}$& 4.1 $\times$ 10$^{-6}$& 1.9 $\times$ 10$^{-5}$&3.0 $\times$ 10$^{43}$ &8.8 $\times$ 10$^{38}$ &      \\ 
Total &0.25 &-- &1.3 & --& 9.9 $\times$ 10$^{-6}$ & 5.3 $\times$ 10$^{-5}$& 6.6$\times$ 10$^{43}$&1.7 $\times$ 10$^{39}$ &\\
\hline
NOMF05-63--1 \\  Red &12 & 17.8 &2.1 $\times$ 10$^{2}$ &3.4 $\times$ 10$^{4}$ &3.5 $\times$ 10$^{-4}$ &6.3 $\times$ 10$^{-3}$ &3.8 $\times$ 10$^{46}$& 1.1 $\times$ 10$^{42}$&   \\ Blue & 4.0&16.0 &6.3 $\times$ 10$^{1}$ & 2.2 $\times$ 10$^{4}$ &1.8 $\times$ 10$^{-4}$& 2.9 $\times$ 10$^{-3}$ &1.0 $\times$ 10$^{46}$ &4.7 $\times$ 10$^{41}$ &\\ 
Total &16 &-- & 2.7 $\times$ 10$^{2}$& -- & 5.3 $\times$ 10$^{-4}$ & 9.2 $\times$ 10$^{-3}$ & 4.8 $\times$ 10$^{46}$ & 1.5 $\times$ 10$^{42}$ &\\
\enddata
\tablecomments{$^a$$i$ = 45$^\circ$ is assumed.}
\label{ta_othrflow_dy}
\end{deluxetable*}

\section{Summary} \label{sec_sum} 
We report the first detection of spatially resolved protostellar outflows and jets in the outer Galaxy with the radio interferometric observations toward Sh 2--283--1a SMM1 ($D_\mathrm{GC}$ = 15.7 kpc and $Z$ = 0.3 $Z_\odot$) with ALMA.
The morphology of the detected outflows and jets as well as their physical and chemical properties were discussed. 
We also reported the new detection of the other 4 outflow sources in the outer Galaxy ($D_\mathrm{GC}$ $\sim$ 17 kpc). 
We summarized the obtained conclusions below.

\begin{enumerate}
    \item{We have detected the outflows and jets with CO($J$ = 3--2) toward Sh 2--283--1a SMM1.     
    The gas velocities relative to the systemic velocity are $\sim$5--50 km s$^{-1}$ for the outflows and $\sim$50--100 km s$^{-1}$ for the jets.
    The morphology of both the outflows and jets are well-collimated and resemble those observed in nearby protostellar systems. 
    The jets show bullet structures, which indicate episodic launching of the jets. 
    We have also detected a jet component with SiO($J$ = 8--7) and the outflow components with HCO$^+$($J$ = 4--3).}

    \item{The physical properties (e.g., mass, momentum, dynamical timescale, etc.) of the outflows and jets are derived from the CO($J$ = 3--2) line.
    According to the derived outflow mass and estimation from $K_s$-band magnitude, the luminosity of Sh 2--283--1a SMM1 would correspond to an intermediate- to high-mass protostellar sources.
    }

    \item{The PV diagram along the flow axis shows the two distinctive structures (Hubble flows and spine-like structures). 
    These spine-like structures would reflect the episodic mass-ejection events with the time intervals of 900--4000 years according to the slopes of the spine-like structures.
    }

    \item{The derived $N\mathrm{(SiO)}$/$N\mathrm{(CO)}$ in the jet bullet based on the LTE analysis is one order of magnitude lower or even less than those observed in nearby star-forming regions. 
    This would indicate the different shock chemistry in the outer Galaxy protostellar core, which may be related to the different dust composition due to its unique low-metallicity environment, although the non-LTE effect introduces significant uncertainty in the derived $N\mathrm{(SiO)}$/$N\mathrm{(CO)}$ ratio as indicated by the RADEX analyses.}
    %

    \item{We also detected the 4 new outflow sources in the outer Galaxy in addition to Sh 2--283--1a SMM1, suggesting that star formation activities are ongoing in the outer Galaxy. }

    \item{The present results would indicate that the earliest star formation processes are not significantly different, at least physically, in a low-metallicity environment of the outer Galaxy. 
    Future follow-up observations are highly required to investigate the smaller scale structures such as protoplanetary disks at low metallicity. }
\end{enumerate}


\begin{acknowledgments}
This paper makes use of the following ALMA data: ADS/JAO.ALMA$\#$2022.1.01270.S. 
ALMA is a partnership of ESO (representing its member states), NSF (USA) and NINS (Japan), together with NRC (Canada), MOST and ASIAA (Taiwan), and KASI (Republic of Korea), in cooperation with the Republic of Chile. 
The Joint ALMA Observatory is operated by ESO, AUI/NRAO and NAOJ. 
This work has made extensive use of the Cologne Database for Molecular Spectroscopy. 
This work was supported by JSPS KAKENHI grant Nos. JP20H05845, JP21H00037, and JP21H01145. 
Finally, we would like to thank an anonymous referee for insightful comments, which substantially improved this paper.
\end{acknowledgments}
%
%


\appendix
\restartappendixnumbering

\section{Information of the target star-forming regions} \label{sec_position} 
\begin{figure*}[tp!]
\begin{center}
\includegraphics[width=18cm]{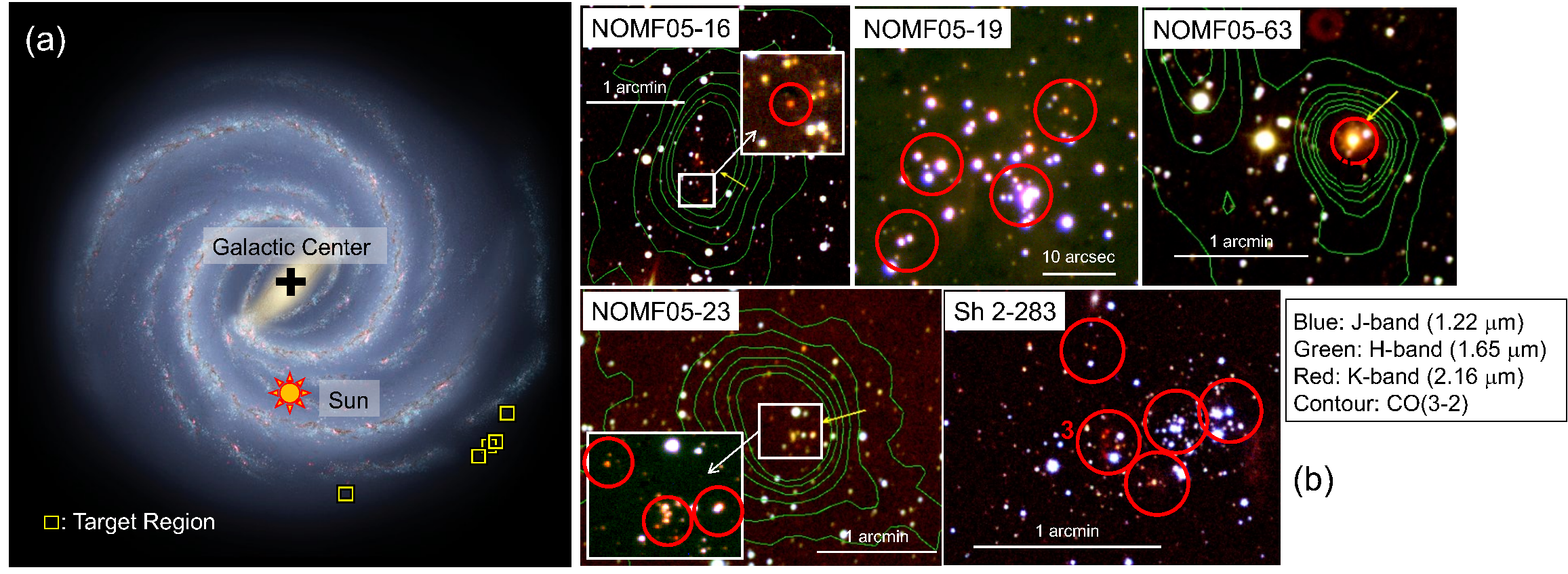}
\caption{The positions of the target star-forming region in the Galaxy and the near-infrared color composite images of the targets are shown.
(a) The target positions are represented as the yellow rectangles.
The background is an artist's conception of the Milky Way (R. Hurt/NASA/JPL-Caltech/ESO).
(b) The positions of the target protostellar candidates are indicated by the red circles.
The number of sources are labeled when multiple sources are enclosed.
Green contours show CO($J$ = 3--2) data for NOMF05-16/23/63 obtained by the ASTE observations. 
}
\label{image_position}
\end{center}
\end{figure*}

Figure \ref{image_position} represents the positions of the target star-forming regions (panel (a)) and the near-infrared color composite images of the targets (panel (b)).
In the panel (b) of this figure, the red circles indicate the positions of the targets, and the green contours show CO($J$ = 3--2) data for NOMF05-16/23/63 obtained by the Atacama Submillimeter Telescope Experiment (ASTE) observations. 

\section{The detected molecular species at the core center} \label{sec_mol} 
\begin{figure*}[tp!]
\begin{center}
\includegraphics[width=12cm]{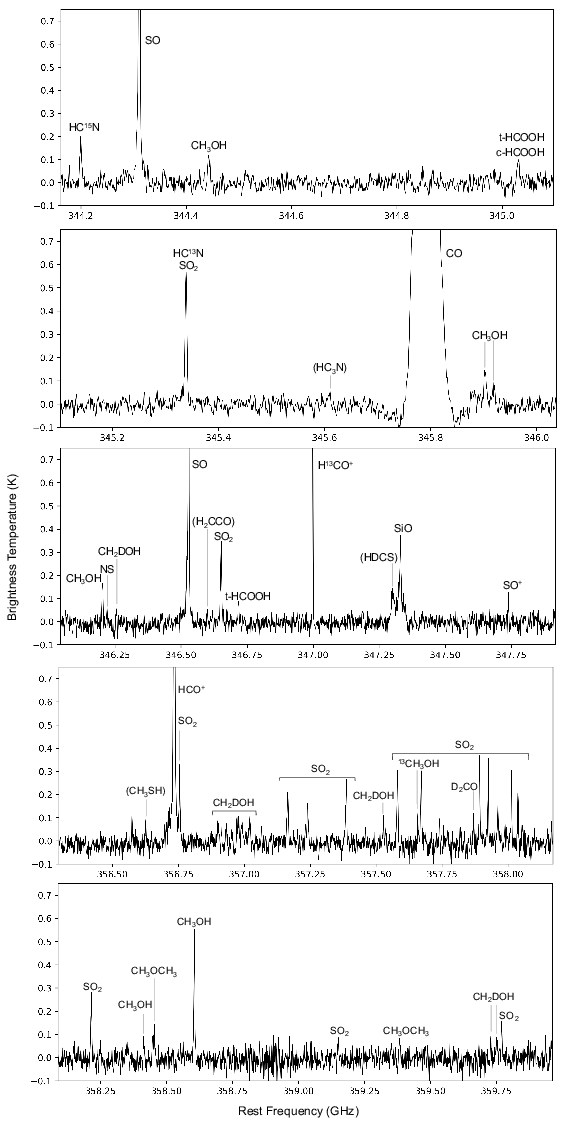}
\caption{
The spectra of Sh 2--283--1a SMM1 which are extracted from the emission peak of 0.87-mm continuum (0$\farcs$89 $\times$ 0$\farcs$70 region). 
Detected emission lines are labeled. 
The parentheses indicate the tentative detection. 
}
\label{image_spec_sum}
\end{center}
\end{figure*}

Figure \ref{image_spec_sum} shows the spectra of Sh 2--283--1a SMM1 which are extracted from the emission peak of 0.87-mm continuum (0$\farcs$89 $\times$ 0$\farcs$70 elliptical region).
In this figure, the parentheses indicate the tentative detection.

\begin{thebibliography}{}
\expandafter\ifx\csname natexlab\endcsname\relax\def\natexlab#1{#1}\fi
\providecommand{\url}[1]{\href{#1}{#1}}
\providecommand{\dodoi}[1]{doi:~\href{http://doi.org/#1}{\nolinkurl{#1}}}
\providecommand{\doeprint}[1]{\href{http://ascl.net/#1}{\nolinkurl{http://ascl.net/#1}}}
\providecommand{\doarXiv}[1]{\href{https://arxiv.org/abs/#1}{\nolinkurl{https://arxiv.org/abs/#1}}}

\bibitem[{{Arce} {et~al.}(2007){Arce}, {Shepherd}, {Gueth}, {Lee}, {Bachiller}, {Rosen}, \& {Beuther}}]{Arce2007}
{Arce}, H.~G., {Shepherd}, D., {Gueth}, F., {et~al.} 2007, in Protostars and Planets V, ed. B.~{Reipurth}, D.~{Jewitt}, \& K.~{Keil}, 245, \dodoi{10.48550/arXiv.astro-ph/0603071}

\bibitem[{{Arellano-C{\'o}rdova} {et~al.}(2021){Arellano-C{\'o}rdova}, {Esteban}, {Garc{\'\i}a-Rojas}, \& {M{\'e}ndez-Delgado}}]{Are21}
{Arellano-C{\'o}rdova}, K.~Z., {Esteban}, C., {Garc{\'\i}a-Rojas}, J., \& {M{\'e}ndez-Delgado}, J.~E. 2021, \mnras, 502, 225, \dodoi{10.1093/mnras/staa3903}

\bibitem[{{Bachiller}(1996)}]{Bac96}
{Bachiller}, R. 1996, \araa, 34, 111, \dodoi{10.1146/annurev.astro.34.1.111}

\bibitem[{{Bachiller} {et~al.}(1990){Bachiller}, {Cernicharo}, {Martin-Pintado}, {Tafalla}, \& {Lazareff}}]{bachiller90}
{Bachiller}, R., {Cernicharo}, J., {Martin-Pintado}, J., {Tafalla}, M., \& {Lazareff}, B. 1990, \aap, 231, 174

\bibitem[{{Balestra} {et~al.}(2007){Balestra}, {Tozzi}, {Ettori}, {Rosati}, {Borgani}, {Mainieri}, {Norman}, \& {Viola}}]{Balestra2007}
{Balestra}, I., {Tozzi}, P., {Ettori}, S., {et~al.} 2007, \aap, 462, 429, \dodoi{10.1051/0004-6361:20065568}

\bibitem[{{Bally}(2016)}]{Bally2016}
{Bally}, J. 2016, \araa, 54, 491, \dodoi{10.1146/annurev-astro-081915-023341}

\bibitem[{{Beuther} {et~al.}(2002){Beuther}, {Schilke}, {Sridharan}, {Menten}, {Walmsley}, \& {Wyrowski}}]{Beuther2002}
{Beuther}, H., {Schilke}, P., {Sridharan}, T.~K., {et~al.} 2002, \aap, 383, 892, \dodoi{10.1051/0004-6361:20011808}

\bibitem[{{Bolatto} {et~al.}(2013){Bolatto}, {Wolfire}, \& {Leroy}}]{Bol13}
{Bolatto}, A.~D., {Wolfire}, M., \& {Leroy}, A.~K. 2013, \araa, 51, 207, \dodoi{10.1146/annurev-astro-082812-140944}

\bibitem[{{Bragan{\c{c}}a} {et~al.}(2019){Bragan{\c{c}}a}, {Daflon}, {Lanz}, {Cunha}, {Bensby}, {McMillan}, {Garmany}, {Glaspey}, {Borges Fernandes}, {Oey}, \& {Hubeny}}]{Bra19}
{Bragan{\c{c}}a}, G.~A., {Daflon}, S., {Lanz}, T., {et~al.} 2019, \aap, 625, A120, \dodoi{10.1051/0004-6361/201834554}

\bibitem[{{CASA Team} {et~al.}(2022){CASA Team}, {Bean}, {Bhatnagar}, {Castro}, {Donovan Meyer}, {Emonts}, {Garcia}, {Garwood}, {Golap}, {Gonzalez Villalba}, {Harris}, {Hayashi}, {Hoskins}, {Hsieh}, {Jagannathan}, {Kawasaki}, {Keimpema}, {Kettenis}, {Lopez}, {Marvil}, {Masters}, {McNichols}, {Mehringer}, {Miel}, {Moellenbrock}, {Montesino}, {Nakazato}, {Ott}, {Petry}, {Pokorny}, {Raba}, {Rau}, {Schiebel}, {Schweighart}, {Sekhar}, {Shimada}, {Small}, {Steeb}, {Sugimoto}, {Suoranta}, {Tsutsumi}, {van Bemmel}, {Verkouter}, {Wells}, {Xiong}, {Szomoru}, {Griffith}, {Glendenning}, \& {Kern}}]{CASA2022}
{CASA Team}, {Bean}, B., {Bhatnagar}, S., {et~al.} 2022, \pasp, 134, 114501, \dodoi{10.1088/1538-3873/ac9642}

\bibitem[{{Cheng} {et~al.}(2019){Cheng}, {Qiu}, {Zhang}, {Wyrowski}, {Menten}, \& {G{\"u}sten}}]{Cheng2019}
{Cheng}, Y., {Qiu}, K., {Zhang}, Q., {et~al.} 2019, \apj, 877, 112, \dodoi{10.3847/1538-4357/ab15d4}

\bibitem[{{Delgado Mena} {et~al.}(2019){Delgado Mena}, {Moya}, {Adibekyan}, {Tsantaki}, {Gonz{\'a}lez Hern{\'a}ndez}, {Israelian}, {Davies}, {Chaplin}, {Sousa}, {Ferreira}, \& {Santos}}]{Delgado2019}
{Delgado Mena}, E., {Moya}, A., {Adibekyan}, V., {et~al.} 2019, \aap, 624, A78, \dodoi{10.1051/0004-6361/201834783}

\bibitem[{{Dunham} {et~al.}(2014){Dunham}, {Arce}, {Mardones}, {Lee}, {Matthews}, {Stutz}, \& {Williams}}]{Dunham2014}
{Dunham}, M.~M., {Arce}, H.~G., {Mardones}, D., {et~al.} 2014, \apj, 783, 29, \dodoi{10.1088/0004-637X/783/1/29}

\bibitem[{{Dutta} {et~al.}(2022){Dutta}, {Lee}, {Hirano}, {Liu}, {Johnstone}, {Liu}, {Tatematsu}, {Goldsmith}, {Sahu}, {Evans}, {Sanhueza}, {Kwon}, {Qin}, {Samal}, {Zhang}, {Kim}, {Shang}, {Lee}, {Moraghan}, {Jhan}, {Li}, {Lee}, {Traficante}, {Juvela}, {Bronfman}, {Eden}, {Soam}, {He}, {Liu}, {Kuan}, {Pelkonen}, {Luo}, {Yi}, \& {Hsu}}]{Dutta2022}
{Dutta}, S., {Lee}, C.-F., {Hirano}, N., {et~al.} 2022, \apj, 931, 130, \dodoi{10.3847/1538-4357/ac67a1}

\bibitem[{{Dutta} {et~al.}(2024){Dutta}, {Lee}, {Johnstone}, {Lee}, {Hirano}, {Di Francesco}, {Moraghan}, {Liu}, {Sahu}, {Liu}, {Tatematsu}, {Goldsmith}, {Lee}, {Li}, {Eden}, {Juvela}, {Bronfman}, {Hsu}, {Kim}, {Kwon}, {Sanhueza}, {Liu}, {L{\'o}pez-V{\'a}zquez}, {Luo}, \& {Yi}}]{Dutta2024}
{Dutta}, S., {Lee}, C.-F., {Johnstone}, D., {et~al.} 2024, \aj, 167, 72, \dodoi{10.3847/1538-3881/ad152b}

\bibitem[{{Fern{\'a}ndez-L{\'o}pez} {et~al.}(2013){Fern{\'a}ndez-L{\'o}pez}, {Girart}, {Curiel}, {Zapata}, {Fonfr{\'\i}a}, \& {Qiu}}]{Fern2013}
{Fern{\'a}ndez-L{\'o}pez}, M., {Girart}, J.~M., {Curiel}, S., {et~al.} 2013, \apj, 778, 72, \dodoi{10.1088/0004-637X/778/1/72}

\bibitem[{{Fern{\'a}ndez-Mart{\'\i}n} {et~al.}(2017){Fern{\'a}ndez-Mart{\'\i}n}, {P{\'e}rez-Montero}, {V{\'\i}lchez}, \& {Mampaso}}]{Fer17}
{Fern{\'a}ndez-Mart{\'\i}n}, A., {P{\'e}rez-Montero}, E., {V{\'\i}lchez}, J.~M., \& {Mampaso}, A. 2017, \aap, 597, A84, \dodoi{10.1051/0004-6361/201628423}

\bibitem[{{Fich} \& {Silkey}(1991)}]{Fich1991}
{Fich}, M., \& {Silkey}, M. 1991, \apj, 366, 107, \dodoi{10.1086/169544}

\bibitem[{{Fischer} {et~al.}(2023){Fischer}, {Hillenbrand}, {Herczeg}, {Johnstone}, {Kospal}, \& {Dunham}}]{Fischer2023}
{Fischer}, W.~J., {Hillenbrand}, L.~A., {Herczeg}, G.~J., {et~al.} 2023, in Astronomical Society of the Pacific Conference Series, Vol. 534, Protostars and Planets VII, ed. S.~{Inutsuka}, Y.~{Aikawa}, T.~{Muto}, K.~{Tomida}, \& M.~{Tamura}, 355, \dodoi{10.48550/arXiv.2203.11257}

\bibitem[{{Fukui} {et~al.}(2015){Fukui}, {Harada}, {Tokuda}, {Morioka}, {Onishi}, {Torii}, {Ohama}, {Hattori}, {Nayak}, {Meixner}, {Sewi{\l}o}, {Indebetouw}, {Kawamura}, {Saigo}, {Yamamoto}, {Tachihara}, {Minamidani}, {Inoue}, {Madden}, {Galametz}, {Lebouteiller}, {Mizuno}, \& {Chen}}]{Fukui2015}
{Fukui}, Y., {Harada}, R., {Tokuda}, K., {et~al.} 2015, \apjl, 807, L4, \dodoi{10.1088/2041-8205/807/1/L4}

\bibitem[{{Gaia Collaboration} {et~al.}(2021){Gaia Collaboration}, {Brown}, {Vallenari}, {Prusti}, {de Bruijne}, {Babusiaux}, {Biermann}, {Creevey}, {Evans}, {Eyer}, {Hutton}, {Jansen}, {Jordi}, {Klioner}, {Lammers}, {Lindegren}, {Luri}, {Mignard}, {Panem}, {Pourbaix}, {Randich}, {Sartoretti}, {Soubiran}, {Walton}, {Arenou}, {Bailer-Jones}, {Bastian}, {Cropper}, {Drimmel}, {Katz}, {Lattanzi}, {van Leeuwen}, {Bakker}, {Cacciari}, {Casta{\~n}eda}, {De Angeli}, {Ducourant}, {Fabricius}, {Fouesneau}, {Fr{\'e}mat}, {Guerra}, {Guerrier}, {Guiraud}, {Jean-Antoine Piccolo}, {Masana}, {Messineo}, {Mowlavi}, {Nicolas}, {Nienartowicz}, {Pailler}, {Panuzzo}, {Riclet}, {Roux}, {Seabroke}, {Sordo}, {Tanga}, {Th{\'e}venin}, {Gracia-Abril}, {Portell}, {Teyssier}, {Altmann}, {Andrae}, {Bellas-Velidis}, {Benson}, {Berthier}, {Blomme}, {Brugaletta}, {Burgess}, {Busso}, {Carry}, {Cellino}, {Cheek}, {Clementini}, {Damerdji}, {Davidson}, {Delchambre}, {Dell'Oro}, {Fern{\'a}ndez-Hern{\'a}ndez}, {Galluccio}, {Garc{\'\i}a-Lario},
  {Garcia-Reinaldos}, {Gonz{\'a}lez-N{\'u}{\~n}ez}, {Gosset}, {Haigron}, {Halbwachs}, {Hambly}, {Harrison}, {Hatzidimitriou}, {Heiter}, {Hern{\'a}ndez}, {Hestroffer}, {Hodgkin}, {Holl}, {Jan{\ss}en}, {Jevardat de Fombelle}, {Jordan}, {Krone-Martins}, {Lanzafame}, {L{\"o}ffler}, {Lorca}, {Manteiga}, {Marchal}, {Marrese}, {Moitinho}, {Mora}, {Muinonen}, {Osborne}, {Pancino}, {Pauwels}, {Petit}, {Recio-Blanco}, {Richards}, {Riello}, {Rimoldini}, {Robin}, {Roegiers}, {Rybizki}, {Sarro}, {Siopis}, {Smith}, {Sozzetti}, {Ulla}, {Utrilla}, {van Leeuwen}, {van Reeven}, {Abbas}, {Abreu Aramburu}, {Accart}, {Aerts}, {Aguado}, {Ajaj}, {Altavilla}, {{\'A}lvarez}, {{\'A}lvarez Cid-Fuentes}, {Alves}, {Anderson}, {Anglada Varela}, {Antoja}, {Audard}, {Baines}, {Baker}, {Balaguer-N{\'u}{\~n}ez}, {Balbinot}, {Balog}, {Barache}, {Barbato}, {Barros}, {Barstow}, {Bartolom{\'e}}, {Bassilana}, {Bauchet}, {Baudesson-Stella}, {Becciani}, {Bellazzini}, {Bernet}, {Bertone}, {Bianchi}, {Blanco-Cuaresma}, {Boch}, {Bombrun}, {Bossini},
  {Bouquillon}, {Bragaglia}, {Bramante}, {Breedt}, {Bressan}, {Brouillet}, {Bucciarelli}, {Burlacu}, {Busonero}, {Butkevich}, {Buzzi}, {Caffau}, {Cancelliere}, {C{\'a}novas}, {Cantat-Gaudin}, {Carballo}, {Carlucci}, {Carnerero}, {Carrasco}, {Casamiquela}, {Castellani}, {Castro-Ginard}, {Castro Sampol}, {Chaoul}, {Charlot}, {Chemin}, {Chiavassa}, {Cioni}, {Comoretto}, {Cooper}, {Cornez}, {Cowell}, {Crifo}, {Crosta}, {Crowley}, {Dafonte}, {Dapergolas}, {David}, {David}, {de Laverny}, {De Luise}, {De March}, {De Ridder}, {de Souza}, {de Teodoro}, {de Torres}, {del Peloso}, {del Pozo}, {Delbo}, {Delgado}, {Delgado}, {Delisle}, {Di Matteo}, {Diakite}, {Diener}, {Distefano}, {Dolding}, {Eappachen}, {Edvardsson}, {Enke}, {Esquej}, {Fabre}, {Fabrizio}, {Faigler}, {Fedorets}, {Fernique}, {Fienga}, {Figueras}, {Fouron}, {Fragkoudi}, {Fraile}, {Franke}, {Gai}, {Garabato}, {Garcia-Gutierrez}, {Garc{\'\i}a-Torres}, {Garofalo}, {Gavras}, {Gerlach}, {Geyer}, {Giacobbe}, {Gilmore}, {Girona}, {Giuffrida}, {Gomel}, {Gomez},
  {Gonzalez-Santamaria}, {Gonz{\'a}lez-Vidal}, {Granvik}, {Guti{\'e}rrez-S{\'a}nchez}, {Guy}, {Hauser}, {Haywood}, {Helmi}, {Hidalgo}, {Hilger}, {H{\l}adczuk}, {Hobbs}, {Holland}, {Huckle}, {Jasniewicz}, {Jonker}, {Juaristi Campillo}, {Julbe}, {Karbevska}, {Kervella}, {Khanna}, {Kochoska}, {Kontizas}, {Kordopatis}, {Korn}, {Kostrzewa-Rutkowska}, {Kruszy{\'n}ska}, {Lambert}, {Lanza}, {Lasne}, {Le Campion}, {Le Fustec}, {Lebreton}, {Lebzelter}, {Leccia}, {Leclerc}, {Lecoeur-Taibi}, {Liao}, {Licata}, {Lindstr{\o}m}, {Lister}, {Livanou}, {Lobel}, {Madrero Pardo}, {Managau}, {Mann}, {Marchant}, {Marconi}, {Marcos Santos}, {Marinoni}, {Marocco}, {Marshall}, {Martin Polo}, {Mart{\'\i}n-Fleitas}, {Masip}, {Massari}, {Mastrobuono-Battisti}, {Mazeh}, {McMillan}, {Messina}, {Michalik}, {Millar}, {Mints}, {Molina}, {Molinaro}, {Moln{\'a}r}, {Montegriffo}, {Mor}, {Morbidelli}, {Morel}, {Morris}, {Mulone}, {Munoz}, {Muraveva}, {Murphy}, {Musella}, {Noval}, {Ord{\'e}novic}, {Orr{\`u}}, {Osinde}, {Pagani}, {Pagano},
  {Palaversa}, {Palicio}, {Panahi}, {Pawlak}, {Pe{\~n}alosa Esteller}, {Penttil{\"a}}, {Piersimoni}, {Pineau}, {Plachy}, {Plum}, {Poggio}, {Poretti}, {Poujoulet}, {Pr{\v{s}}a}, {Pulone}, {Racero}, {Ragaini}, {Rainer}, {Raiteri}, {Rambaux}, {Ramos}, {Ramos-Lerate}, {Re Fiorentin}, {Regibo}, {Reyl{\'e}}, {Ripepi}, {Riva}, {Rixon}, {Robichon}, {Robin}, {Roelens}, {Rohrbasser}, {Romero-G{\'o}mez}, {Rowell}, {Royer}, {Rybicki}, {Sadowski}, {Sagrist{\`a} Sell{\'e}s}, {Sahlmann}, {Salgado}, {Salguero}, {Samaras}, {Sanchez Gimenez}, {Sanna}, {Santove{\~n}a}, {Sarasso}, {Schultheis}, {Sciacca}, {Segol}, {Segovia}, {S{\'e}gransan}, {Semeux}, {Shahaf}, {Siddiqui}, {Siebert}, {Siltala}, {Slezak}, {Smart}, {Solano}, {Solitro}, {Souami}, {Souchay}, {Spagna}, {Spoto}, {Steele}, {Steidelm{\"u}ller}, {Stephenson}, {S{\"u}veges}, {Szabados}, {Szegedi-Elek}, {Taris}, {Tauran}, {Taylor}, {Teixeira}, {Thuillot}, {Tonello}, {Torra}, {Torra}, {Turon}, {Unger}, {Vaillant}, {van Dillen}, {Vanel}, {Vecchiato}, {Viala}, {Vicente},
  {Voutsinas}, {Weiler}, {Wevers}, {Wyrzykowski}, {Yoldas}, {Yvard}, {Zhao}, {Zorec}, {Zucker}, {Zurbach}, \& {Zwitter}}]{Gaia2021}
{Gaia Collaboration}, {Brown}, A.~G.~A., {Vallenari}, A., {et~al.} 2021, \aap, 649, A1, \dodoi{10.1051/0004-6361/202039657}

\bibitem[{{Ginsburg} {et~al.}(2011){Ginsburg}, {Bally}, \& {Williams}}]{Ginsburg2011}
{Ginsburg}, A., {Bally}, J., \& {Williams}, J.~P. 2011, \mnras, 418, 2121, \dodoi{10.1111/j.1365-2966.2011.19279.x}

\bibitem[{{Hamedani Golshan} {et~al.}(2024){Hamedani Golshan}, {S{\'a}nchez-Monge}, {Schilke}, {Sewi{\l}o}, {M{\"o}ller}, {Veena}, \& {Fuller}}]{Gol24}
{Hamedani Golshan}, R., {S{\'a}nchez-Monge}, {\'A}., {Schilke}, P., {et~al.} 2024, arXiv e-prints, arXiv:2405.01710, \dodoi{10.48550/arXiv.2405.01710}

\bibitem[{{Hirano} {et~al.}(2010){Hirano}, {Ho}, {Liu}, {Shang}, {Lee}, \& {Bourke}}]{Hirano2010}
{Hirano}, N., {Ho}, P. P.~T., {Liu}, S.-Y., {et~al.} 2010, \apj, 717, 58, \dodoi{10.1088/0004-637X/717/1/58}

\bibitem[{{Izumi} {et~al.}(2024){Izumi}, {Ressler}, {Lau}, {Koch}, {Saito}, {Kobayashi}, \& {Yasui}}]{Izumi2024}
{Izumi}, N., {Ressler}, M.~E., {Lau}, R.~M., {et~al.} 2024, \aj, 168, 68, \dodoi{10.3847/1538-3881/ad4e2e}

\bibitem[{{Lee} {et~al.}(2000){Lee}, {Mundy}, {Reipurth}, {Ostriker}, \& {Stone}}]{Lee2000}
{Lee}, C.-F., {Mundy}, L.~G., {Reipurth}, B., {Ostriker}, E.~C., \& {Stone}, J.~M. 2000, \apj, 542, 925, \dodoi{10.1086/317056}

\bibitem[{{Lefloch} {et~al.}(2015){Lefloch}, {Gusdorf}, {Codella}, {Eisl{\"o}ffel}, {Neri}, {G{\'o}mez-Ruiz}, {G{\"u}sten}, {Leurini}, {Risacher}, \& {Benedettini}}]{Lefloch2015}
{Lefloch}, B., {Gusdorf}, A., {Codella}, C., {et~al.} 2015, \aap, 581, A4, \dodoi{10.1051/0004-6361/201425521}

\bibitem[{{Lindegren} {et~al.}(2021){Lindegren}, {Bastian}, {Biermann}, {Bombrun}, {de Torres}, {Gerlach}, {Geyer}, {Hern{\'a}ndez}, {Hilger}, {Hobbs}, {Klioner}, {Lammers}, {McMillan}, {Ramos-Lerate}, {Steidelm{\"u}ller}, {Stephenson}, \& {van Leeuwen}}]{Lindegren2021}
{Lindegren}, L., {Bastian}, U., {Biermann}, M., {et~al.} 2021, \aap, 649, A4, \dodoi{10.1051/0004-6361/202039653}

\bibitem[{{Louvet} {et~al.}(2018){Louvet}, {Dougados}, {Cabrit}, {Mardones}, {M{\'e}nard}, {Tabone}, {Pinte}, \& {Dent}}]{Louvet2018}
{Louvet}, F., {Dougados}, C., {Cabrit}, S., {et~al.} 2018, \aap, 618, A120, \dodoi{10.1051/0004-6361/201731733}

\bibitem[{{Lucas} {et~al.}(2008){Lucas}, {Hoare}, {Longmore}, {Schr{\"o}der}, {Davis}, {Adamson}, {Bandyopadhyay}, {de Grijs}, {Smith}, {Gosling}, {Mitchison}, {G{\'a}sp{\'a}r}, {Coe}, {Tamura}, {Parker}, {Irwin}, {Hambly}, {Bryant}, {Collins}, {Cross}, {Evans}, {Gonzalez-Solares}, {Hodgkin}, {Lewis}, {Read}, {Riello}, {Sutorius}, {Lawrence}, {Drew}, {Dye}, \& {Thompson}}]{Lucas2008}
{Lucas}, P.~W., {Hoare}, M.~G., {Longmore}, A., {et~al.} 2008, \mnras, 391, 136, \dodoi{10.1111/j.1365-2966.2008.13924.x}

\bibitem[{{Machida} \& {Basu}(2019)}]{Machida2019}
{Machida}, M.~N., \& {Basu}, S. 2019, \apj, 876, 149, \dodoi{10.3847/1538-4357/ab18a7}

\bibitem[{{Mangum} \& {Shirley}(2015)}]{Mangum2015}
{Mangum}, J.~G., \& {Shirley}, Y.~L. 2015, \pasp, 127, 266, \dodoi{10.1086/680323}

\bibitem[{{Matsushita} {et~al.}(2019){Matsushita}, {Takahashi}, {Machida}, \& {Tomisaka}}]{Matsushita2019}
{Matsushita}, Y., {Takahashi}, S., {Machida}, M.~N., \& {Tomisaka}, K. 2019, \apj, 871, 221, \dodoi{10.3847/1538-4357/aaf1b6}

\bibitem[{{Maud} {et~al.}(2015){Maud}, {Moore}, {Lumsden}, {Mottram}, {Urquhart}, \& {Hoare}}]{Maud2015}
{Maud}, L.~T., {Moore}, T.~J.~T., {Lumsden}, S.~L., {et~al.} 2015, \mnras, 453, 645, \dodoi{10.1093/mnras/stv1635}

\bibitem[{{M{\"u}ller} {et~al.}(2005){M{\"u}ller}, {Schl{\"o}der}, {Stutzki}, {Schlemmer}, {Giesen}, \& {Schilke}}]{Muller2005}
{M{\"u}ller}, H. S.~P., {Schl{\"o}der}, F., {Stutzki}, J., {et~al.} 2005, in Astrochemistry: Recent Successes and Current Challenges, ed. D.~C. {Lis}, G.~A. {Blake}, \& E.~{Herbst}, Vol. 231, 62

\bibitem[{{M{\"u}ller} {et~al.}(2001){M{\"u}ller}, {Thorwirth}, {Roth}, \& {Winnewisser}}]{Muller2001}
{M{\"u}ller}, H.~S.~P., {Thorwirth}, S., {Roth}, D.~A., \& {Winnewisser}, G. 2001, \aap, 370, L49, \dodoi{10.1051/0004-6361:20010367}

\bibitem[{{Nakagawa} {et~al.}(2005){Nakagawa}, {Onishi}, {Mizuno}, \& {Fukui}}]{Nakagawa2005}
{Nakagawa}, M., {Onishi}, T., {Mizuno}, A., \& {Fukui}, Y. 2005, \pasj, 57, 917, \dodoi{10.1093/pasj/57.6.917}

\bibitem[{{Nisini} {et~al.}(2007){Nisini}, {Codella}, {Giannini}, {Santiago Garcia}, {Richer}, {Bachiller}, \& {Tafalla}}]{Nisini2007}
{Nisini}, B., {Codella}, C., {Giannini}, T., {et~al.} 2007, \aap, 462, 163, \dodoi{10.1051/0004-6361:20065621}

\bibitem[{{Nony} {et~al.}(2020){Nony}, {Motte}, {Louvet}, {Plunkett}, {Gusdorf}, {Fechtenbaum}, {Pouteau}, {Lefloch}, {Bontemps}, {Molet}, \& {Robitaille}}]{Nony2020}
{Nony}, T., {Motte}, F., {Louvet}, F., {et~al.} 2020, \aap, 636, A38, \dodoi{10.1051/0004-6361/201937046}

\bibitem[{{Omura} {et~al.}(2024){Omura}, {Tokuda}, \& {Machida}}]{Omura2024}
{Omura}, M., {Tokuda}, K., \& {Machida}, M.~N. 2024, arXiv e-prints, arXiv:2401.03086, \dodoi{10.48550/arXiv.2401.03086}

\bibitem[{{Plunkett} {et~al.}(2015){Plunkett}, {Arce}, {Mardones}, {van Dokkum}, {Dunham}, {Fern{\'a}ndez-L{\'o}pez}, {Gallardo}, \& {Corder}}]{Plunkett2015}
{Plunkett}, A.~L., {Arce}, H.~G., {Mardones}, D., {et~al.} 2015, \nat, 527, 70, \dodoi{10.1038/nature15702}

\bibitem[{{Pyo} {et~al.}(2024){Pyo}, {Hayashi}, {Takami}, \& {Beck}}]{Pyo2024}
{Pyo}, T.-S., {Hayashi}, M., {Takami}, M., \& {Beck}, T.~L. 2024, \apj, 963, 159, \dodoi{10.3847/1538-4357/ad1f59}

\bibitem[{{Qiu} \& {Zhang}(2009)}]{Qiu2009}
{Qiu}, K., \& {Zhang}, Q. 2009, \apjl, 702, L66, \dodoi{10.1088/0004-637X/702/1/L66}

\bibitem[{{Rafelski} {et~al.}(2012){Rafelski}, {Wolfe}, {Prochaska}, {Neeleman}, \& {Mendez}}]{Rafelski2012}
{Rafelski}, M., {Wolfe}, A.~M., {Prochaska}, J.~X., {Neeleman}, M., \& {Mendez}, A.~J. 2012, \apj, 755, 89, \dodoi{10.1088/0004-637X/755/2/89}

\bibitem[{{Reid} {et~al.}(2014){Reid}, {Menten}, {Brunthaler}, {Zheng}, {Dame}, {Xu}, {Wu}, {Zhang}, {Sanna}, {Sato}, {Hachisuka}, {Choi}, {Immer}, {Moscadelli}, {Rygl}, \& {Bartkiewicz}}]{Reid2014}
{Reid}, M.~J., {Menten}, K.~M., {Brunthaler}, A., {et~al.} 2014, \apj, 783, 130, \dodoi{10.1088/0004-637X/783/2/130}

\bibitem[{{Robitaille} {et~al.}(2006){Robitaille}, {Whitney}, {Indebetouw}, {Wood}, \& {Denzmore}}]{Rob06}
{Robitaille}, T.~P., {Whitney}, B.~A., {Indebetouw}, R., {Wood}, K., \& {Denzmore}, P. 2006, \apjs, 167, 256, \dodoi{10.1086/508424}

\bibitem[{{Santiago-Garc{\'\i}a} {et~al.}(2009){Santiago-Garc{\'\i}a}, {Tafalla}, {Johnstone}, \& {Bachiller}}]{Santiago2009}
{Santiago-Garc{\'\i}a}, J., {Tafalla}, M., {Johnstone}, D., \& {Bachiller}, R. 2009, \aap, 495, 169, \dodoi{10.1051/0004-6361:200810739}

\bibitem[{{Schilke} {et~al.}(1997){Schilke}, {Walmsley}, {Pineau des Forets}, \& {Flower}}]{Sch97_SiO}
{Schilke}, P., {Walmsley}, C.~M., {Pineau des Forets}, G., \& {Flower}, D.~R. 1997, \aap, 321, 293

\bibitem[{{Sewi{\l}o} {et~al.}(2018){Sewi{\l}o}, {Indebetouw}, {Charnley}, {Zahorecz}, {Oliveira}, {van Loon}, {Ward}, {Chen}, {Wiseman}, {Fukui}, {Kawamura}, {Meixner}, {Onishi}, \& {Schilke}}]{Sewilo2018}
{Sewi{\l}o}, M., {Indebetouw}, R., {Charnley}, S.~B., {et~al.} 2018, \apjl, 853, L19, \dodoi{10.3847/2041-8213/aaa079}

\bibitem[{{Sewi{\l}o} {et~al.}(2022){Sewi{\l}o}, {Cordiner}, {Charnley}, {Oliveira}, {Garcia-Berrios}, {Schilke}, {Ward}, {Wiseman}, {Indebetouw}, {Tokuda}, {van Loon}, {S{\'a}nchez-Monge}, {Allen}, {Chen}, {Hamedani Golshan}, {Karska}, {Kristensen}, {Kurtz}, {M{\"o}ller}, {Onishi}, \& {Zahorecz}}]{Sewilo2022}
{Sewi{\l}o}, M., {Cordiner}, M., {Charnley}, S.~B., {et~al.} 2022, \apj, 931, 102, \dodoi{10.3847/1538-4357/ac4e8f}

\bibitem[{{Shimonishi} {et~al.}(2020){Shimonishi}, {Das}, {Sakai}, {Tanaka}, {Aikawa}, {Onaka}, {Watanabe}, \& {Nishimura}}]{Shimonishi2020}
{Shimonishi}, T., {Das}, A., {Sakai}, N., {et~al.} 2020, \apj, 891, 164, \dodoi{10.3847/1538-4357/ab6e6b}

\bibitem[{{Shimonishi} {et~al.}(2021){Shimonishi}, {Izumi}, {Furuya}, \& {Yasui}}]{Shimonishi2021}
{Shimonishi}, T., {Izumi}, N., {Furuya}, K., \& {Yasui}, C. 2021, \apj, 922, 206, \dodoi{10.3847/1538-4357/ac289b}

\bibitem[{{Shimonishi} {et~al.}(2016){Shimonishi}, {Onaka}, {Kawamura}, \& {Aikawa}}]{Shimonishi2016}
{Shimonishi}, T., {Onaka}, T., {Kawamura}, A., \& {Aikawa}, Y. 2016, \apj, 827, 72, \dodoi{10.3847/0004-637X/827/1/72}

\bibitem[{{Shimonishi} {et~al.}(2023){Shimonishi}, {Tanaka}, {Zhang}, \& {Furuya}}]{Shimonishi2023}
{Shimonishi}, T., {Tanaka}, K. E.~I., {Zhang}, Y., \& {Furuya}, K. 2023, \apjl, 946, L41, \dodoi{10.3847/2041-8213/acc031}

\bibitem[{{Snell} {et~al.}(1980){Snell}, {Loren}, \& {Plambeck}}]{snell1980}
{Snell}, R.~L., {Loren}, R.~B., \& {Plambeck}, R.~L. 1980, \apjl, 239, L17, \dodoi{10.1086/183283}

\bibitem[{{Tabone} {et~al.}(2020){Tabone}, {Godard}, {Pineau des For{\^e}ts}, {Cabrit}, \& {van Dishoeck}}]{Tab20}
{Tabone}, B., {Godard}, B., {Pineau des For{\^e}ts}, G., {Cabrit}, S., \& {van Dishoeck}, E.~F. 2020, \aap, 636, A60, \dodoi{10.1051/0004-6361/201937383}

\bibitem[{{Tafalla} {et~al.}(2010){Tafalla}, {Santiago-Garc{\'\i}a}, {Hacar}, \& {Bachiller}}]{Tafalla2010}
{Tafalla}, M., {Santiago-Garc{\'\i}a}, J., {Hacar}, A., \& {Bachiller}, R. 2010, \aap, 522, A91, \dodoi{10.1051/0004-6361/201015158}

\bibitem[{{Takahashi} {et~al.}(2024){Takahashi}, {Machida}, {Omura}, {Johnstone}, {Saigo}, {Harada}, {Tomisaka}, {Ho}, {Zapata}, {Mairs}, {Herczeg}, {Taniguchi}, {Liu}, \& {Sato}}]{Takahashi2024}
{Takahashi}, S., {Machida}, M.~N., {Omura}, M., {et~al.} 2024, arXiv e-prints, arXiv:2401.13204, \dodoi{10.48550/arXiv.2401.13204}

\bibitem[{{Takami} {et~al.}(2023){Takami}, {G{\"u}nther}, {Schneider}, {Beck}, {Karr}, {Ohyama}, {Galv{\'a}n-Madrid}, {Uyama}, {White}, {Grankin}, {Coffey}, {Liu}, {Fukagawa}, {Manset}, {Chen}, {Pyo}, {Shang}, {Ray}, {Otsuka}, \& {Chou}}]{Takami2023}
{Takami}, M., {G{\"u}nther}, H.~M., {Schneider}, P.~C., {et~al.} 2023, \apjs, 264, 1, \dodoi{10.3847/1538-4365/ac9afc}

\bibitem[{{Tokuda} {et~al.}(2022){Tokuda}, {Zahorecz}, {Kunitoshi}, {Higashino}, {Tanaka}, {Konishi}, {Suzuki}, {Kitano}, {Harada}, {Shimonishi}, {Neelamkodan}, {Fukui}, {Kawamura}, {Onishi}, \& {Machida}}]{Tokuda2022}
{Tokuda}, K., {Zahorecz}, S., {Kunitoshi}, Y., {et~al.} 2022, \apjl, 936, L6, \dodoi{10.3847/2041-8213/ac81c1}

\bibitem[{{Tokuda} {et~al.}(2025){Tokuda}, {Kunitoshi}, {Zahorecz}, {Tanaka}, {Murakoso}, {Harada}, {Kobayashi}, {Inoue}, {Sewi{\l}o}, {Konishi}, {Shimonishi}, {Zhang}, {Fukui}, {Kawamura}, {Onishi}, \& {Machida}}]{Tok25}
{Tokuda}, K., {Kunitoshi}, Y., {Zahorecz}, S., {et~al.} 2025, arXiv e-prints, arXiv:2501.02190, \dodoi{10.48550/arXiv.2501.02190}

\bibitem[{{Torrelles} {et~al.}(2011){Torrelles}, {Patel}, {Curiel}, {Estalella}, {G{\'o}mez}, {Rodr{\'\i}guez}, {Cant{\'o}}, {Anglada}, {Vlemmings}, {Garay}, {Raga}, \& {Ho}}]{Torrelles2011}
{Torrelles}, J.~M., {Patel}, N.~A., {Curiel}, S., {et~al.} 2011, \mnras, 410, 627, \dodoi{10.1111/j.1365-2966.2010.17483.x}

\bibitem[{{Tychoniec} {et~al.}(2019){Tychoniec}, {Hull}, {Kristensen}, {Tobin}, {Le Gouellec}, \& {van Dishoeck}}]{Łukasz2019}
{Tychoniec}, {\L}., {Hull}, C. L.~H., {Kristensen}, L.~E., {et~al.} 2019, \aap, 632, A101, \dodoi{10.1051/0004-6361/201935409}

\bibitem[{{van der Tak} {et~al.}(2007){van der Tak}, {Black}, {Sch{\"o}ier}, {Jansen}, \& {van Dishoeck}}]{vdT07}
{van der Tak}, F.~F.~S., {Black}, J.~H., {Sch{\"o}ier}, F.~L., {Jansen}, D.~J., \& {van Dishoeck}, E.~F. 2007, \aap, 468, 627, \dodoi{10.1051/0004-6361:20066820}

\bibitem[{{Wright} {et~al.}(2010){Wright}, {Eisenhardt}, {Mainzer}, {Ressler}, {Cutri}, {Jarrett}, {Kirkpatrick}, {Padgett}, {McMillan}, {Skrutskie}, {Stanford}, {Cohen}, {Walker}, {Mather}, {Leisawitz}, {Gautier}, {McLean}, {Benford}, {Lonsdale}, {Blain}, {Mendez}, {Irace}, {Duval}, {Liu}, {Royer}, {Heinrichsen}, {Howard}, {Shannon}, {Kendall}, {Walsh}, {Larsen}, {Cardon}, {Schick}, {Schwalm}, {Abid}, {Fabinsky}, {Naes}, \& {Tsai}}]{Wri10}
{Wright}, E.~L., {Eisenhardt}, P.~R.~M., {Mainzer}, A.~K., {et~al.} 2010, \aj, 140, 1868, \dodoi{10.1088/0004-6256/140/6/1868}

\bibitem[{{Wu} {et~al.}(2004{\natexlab{a}}){Wu}, {Wei}, {Zhao}, {Shi}, {Yu}, {Qin}, \& {Huang}}]{Wu2004}
{Wu}, Y., {Wei}, Y., {Zhao}, M., {et~al.} 2004{\natexlab{a}}, \aap, 426, 503, \dodoi{10.1051/0004-6361:20035767}

\bibitem[{{Wu} {et~al.}(2004{\natexlab{b}}){Wu}, {Wei}, {Zhao}, {Shi}, {Yu}, {Qin}, \& {Huang}}]{Wu04}
---. 2004{\natexlab{b}}, \aap, 426, 503, \dodoi{10.1051/0004-6361:20035767}

\bibitem[{{Yamamura} {et~al.}(2010){Yamamura}, {Makiuti}, {Ikeda}, {Fukuda}, {Oyabu}, {Koga}, \& {White}}]{Yam10}
{Yamamura}, I., {Makiuti}, S., {Ikeda}, N., {et~al.} 2010, VizieR Online Data Catalog, II/298

\bibitem[{{Zinchenko} {et~al.}(2021){Zinchenko}, {Dewangan}, {Baug}, {Ojha}, \& {Bhadari}}]{Zinchenko2021}
{Zinchenko}, I.~I., {Dewangan}, L.~K., {Baug}, T., {Ojha}, D.~K., \& {Bhadari}, N.~K. 2021, \mnras, 506, L45, \dodoi{10.1093/mnrasl/slab070}

\bibitem[{{Zinchenko} {et~al.}(2024){Zinchenko}, {Liu}, \& {Su}}]{Zinchenko2024}
{Zinchenko}, I.~I., {Liu}, S.~Y., \& {Su}, Y.~N. 2024, arXiv e-prints, arXiv:2411.00116.
\newblock \doarXiv{2411.00116}

\bibitem[{{Zinchenko} {et~al.}(2020){Zinchenko}, {Liu}, {Su}, {Wang}, \& {Wang}}]{Zinchenko2020}
{Zinchenko}, I.~I., {Liu}, S.-Y., {Su}, Y.-N., {Wang}, K.-S., \& {Wang}, Y. 2020, \apj, 889, 43, \dodoi{10.3847/1538-4357/ab5c18}

\end{thebibliography}

\end{document}